\documentclass[pss]{wiley2sp} 
\usepackage{amsmath,amssymb}
\usepackage{bm}              

\usepackage[colorlinks]{hyperref}
\usepackage{cite}
\newcommand{\rot}{\mathop{\rm rot}\nolimits}
\newcommand{\erf}{\mathop{\rm erf}\nolimits}

\begin{document}

\title{Spin and valley dynamics of excitons in transition metal dichalcogenide monolayers}

\titlerunning{Spin and valley dynamics of excitons in transition metal dichalcogenide monolayers}

\author{%
  M.M. Glazov\textsuperscript{\Ast,\textsf{\bfseries 1}},
  E.L. Ivchenko\textsuperscript{\textsf{\bfseries 1}},
  G. Wang\textsuperscript{\textsf{\bfseries 2}},
  T. Amand\textsuperscript{\textsf{\bfseries 2}}, 
  X. Marie\textsuperscript{\textsf{\bfseries 2}}, 
  B. Urbaszek\textsuperscript{\textsf{\bfseries 2}}, 
   B.L. Liu\textsuperscript{\textsf{\bfseries 3}} }

\authorrunning{M.M. Glazov et al.}

\mail{e-mail
  \textsf{glazov@coherent.ioffe.ru}, Phone:
  +7-911-913-0436, Fax: +7-812-297-1017}

\institute{%
  \textsuperscript{1}\,Ioffe Institute, 194021, St.-Petersburg, Russia\\
  \textsuperscript{2}\,Universit\'{e} de Toulouse, INSA-CNRS-UPS, LPCNO, 31077 Toulouse, France\\
  \textsuperscript{3}\,Institute of Physics, Chinese Academy of Sciences, 100190 Beijing, People�s Republic of China}

\received{XXXX, revised XXXX, accepted XXXX} 
\published{XXXX} 


\abstract{%
%
%
%
\abstcol{%
Monolayers of transition metal dichalcogenides, namely, molybdenum and tungsten disulfides and diselenides demonstrate unusual optical properties related to the spin-valley locking effect. Particularly, excitation of monolayers by circularly polarized light selectively creates electron-hole pairs or excitons in non-equivalent valleys in momentum space, depending on the light helicity. This allows studying the inter-valley  dynamics of charge carriers and Coulomb complexes by means of optical spectroscopy. Here we present a concise review of the neutral exciton fine structure and its spin and valley dynamics in monolayers of transition metal dichalcogenides. It is demonstrated that the long-range exchange interaction between an electron and a hole in the exciton is an efficient mechanism for rapid mixing between bright excitons made of electron-hole pairs in different valleys. We discuss the physical origin of the long-range exchange interaction and outline its derivation in both the electrodynamical and $\bm k \cdot \bm p$ approaches. We further present a model of bright exciton spin dynamics driven by an interplay between the long-range exchange interaction and  scattering. Finally, we discuss the application of the model to describe recent experimental data obtained by time-resolved photoluminescence and Kerr rotation techniques.
  }  }

\titlefigure[width=7.9cm,height=5.7cm]{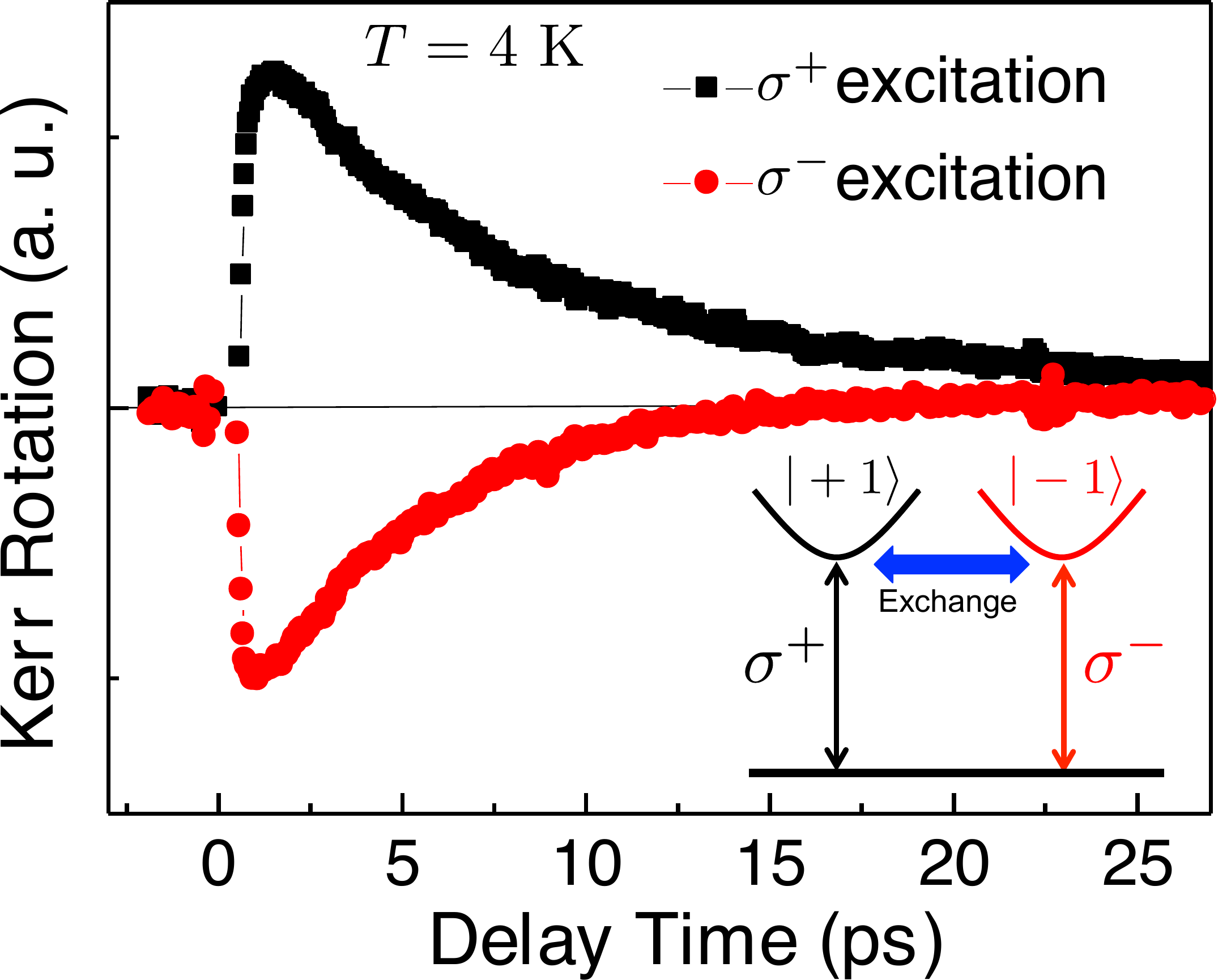}
\titlefigurecaption{%
  Kerr rotation dynamics at $T = 4$~K for a $\sigma^+$ and $\sigma^-$ pump beam in WSe$_2$. Inset: Schematics of the optical selection rules of the excitons photogenerated from charge carriers in $\bm K_\pm$ valleys and their coupling induced by the long-range exchange interaction. From [C. R. Zhu, K. Zhang, M. Glazov, B. Urbaszek, T. Amand, Z. W. Ji, B. L. Liu, and X. Marie, Exciton valley dynamics probed by Kerr rotation in WSe2 monolayers, Phys. Rev. B {\bf 90}, 161302(R) (2014)]. }

\maketitle   

\section{Introduction}\label{sec:intro}

Transition metal dichalcogenide monolayers have remarkable electronic, optical, optoelectronic and chemical properties~\cite{novoselov05a,Wang:2012aa,doi:10.1021/cr300263a}. They can serve as a building blocks for novel type of heterostructures, known as van der Waals heterostructures, where layers of different materials can be isolated and then assembled in a tailored sequence~\cite{Geim:2013aa}. Monolayers of MX$_2$ where M=Mo,W stands for the transition metal and X=S,Se,Te stands for the chalcogen have a hexagonal lattice like graphene. In contrast to graphene, where conduction and valence bands merge at the edges of the Brillouin zone~\cite{Geim2007}, they demonstrate direct bandgaps $\sim 2$~eV realized at the edges, points $\bm K_\pm$, of the Brillouin zone~\cite{doi:10.1021/cr300263a}. 

Strong spin-orbit interaction in MX$_2$ monolayers leads to prominent effects termed as \emph{spin-valley locking}~\cite{Xu:2014cr}. First, valence and conduction band states are spin split already at the $\bm K_\pm$ points~\cite{Xiao:2012cr,Kormanyos:2013dq,Liu:2013if,PhysRevB.88.245436,PhysRevX.4.011034}. The splitting is of opposite signs in the $\bm K_+$ and $\bm K_-$ valleys and acts as a valley-dependent static magnetic field applied perpendicular to the monolayer plane. Second, optical transition rules for interband transitions in the valleys $\bm K_\pm$ are chiral: $\sigma^+$ light induces transitions in one valley of the energy spectrum, $\bm K_+$, while $\sigma^-$ light induces transitions in the other one, $\bm K_-$~\cite{Cao:2012fk,Mak:2012qf,Zeng:2012ys}. Hence, excitation of MX$_2$ samples by circularly polarized light induces both spin and valley polarization of charge carriers. These  phenomena make transition metal dichalcogenides ideal systems for studying spin and valley dynamics in both experiment and theory, see for example Refs.~\cite{Xu:2014cr,Cao:2012fk,Mak:2012qf,Zeng:2012ys,Jones:2013tg,Wang:2013hb,Mak27062014}.

Substantial optical orientation in transition metal dichalcogenides such as MoS$_2$ and WSe$_2$ related to selective excitation of the valleys by circularly polarized light has been indeed revealed in experiments, e.g.~\cite{Sallen:2012qf,Kioseoglou,PhysRevLett.112.047401,Zhu:2013ve,doi:10.1021/nl403742j,Wang:2013hb,PhysRevB.90.161302,2015arXiv150207088Y}. Moreover, single particle spin states and valley coherence were expected to be extremely robust for several interlinked reasons: First, it is necessary to transfer a huge wavevector, on the order of inverse lattice constant, for the charge carrier to change valley. Second, the carrier needs also to flip its spin, which is energetically unfavourable due to the large spin-orbit splittings~\cite{Li:2013qf,Molina-Sanchez:2013mi,Ochoa:2013wd,MWWuMoS2,Song:2013uq}. Despite the valley index stability expected in this single particle picture, recent  optical spectroscopy experiments demonstrate surprisingly rapid intervalley transfer on a picosecond time scale~\cite{doi:10.1021/nl403742j,Wang:2013hb,PhysRevB.90.161302,2015arXiv150207088Y}.

In reality the optical properties of MX$_2$ monolayers are governed not by free carriers but by excitons, electron-hole pairs strongly bound by Coulomb attraction. In this feature article we will show in detail why and how excitons govern the polarization dynamics observed in optical spectroscopy. This allows understanding how the optically generated spin/valley polarization can decay on a picosecond timescale, in good agreement with the experimental findings~\cite{glazov2014exciton}. Indeed, weak screening of the Coulomb interaction and relatively large effective masses of electrons and holes result in strong, tightly-bound, excitons in transition metal dichalcogenides monolayers. The exciton binding energies in these materials $E_b\sim 0.5$~eV~\cite{Cheiwchanchamnangij:2012pi,Mak:2013lh,Chernikov:2014a,2014arXiv1404.0056W,PhysRevB.91.075310} exceed by far those in conventional two-dimensional systems based GaAs-like semiconductors where $E_b \lesssim 10^{-2}$~eV~\cite{Vinattiery,Dareys1993353,maialle93}. Experiments and theoretical modeling reveal also  photoluminescence lifetimes in the picosecond range indicating high exciton oscillator strengths of excitons in MX$_2$ monolayers~\cite{PhysRevLett.112.047401,KornMoS2,PhysRevB.90.075413,doi:10.1021/nl503799t}.
Despite strong suppression of individual carrier spin flips, the exchange interaction in the electron-hole pair forming an exciton~\cite{glazov2014exciton,PhysRevB.89.205303,Yu:2014fk-1} turns out to be strong enough to enable simultaneous spin flip of an electron and a hole~\cite{BAP}. The exciton spin dynamics in monolayers of transition metal dichalcogenides is, thus, similar to that of excitons in conventional two-dimensional semiconductor structures~\cite{maialle93}.

Here we provide a brief overview of exciton fine structure and spin dynamics in monolayers MX$_2$. The article is organized as follows: We start in Sec.~\ref{sec:band} with the basic introduction to the band structure of transition metal dichalcogenides monolayers and to excitonic effects in these materials. A brief overview of exciton complexes investigated in theory and experiment in MX$_2$ monolayers is given in Sec.~\ref{sec:exc}. Further we discuss the fine structure of the bright excitons and review theoretical approaches to calculate the exchange interaction between an electron and a hole in Sec.~\ref{sec:fine}. Next, in Sec.~\ref{sec:theor} we address theoretically exciton spin dynamics governed by the exchange interaction in MX$_2$ monolayers and discuss recent experimental findings on excitonic spin dynamics in Sec.~\ref{sec:exper}. The outlook and conclusions are presented in Sec.~\ref{sec:out} 

\section{Band structure}\label{sec:band}

The band structure of transition metal dichalcogenide monolayers is studied theoretically by various methods, see Ref.~\cite{C4CS00301B} for a review, including both first principles approaches, e.g.~\cite{Cheiwchanchamnangij:2012pi,Molina-Sanchez:2013mi,ANDP:ANDP201400128,PhysRevB.88.245436,PhysRevB.87.155304}, and empirical ones. The latter involve empirical tight-binding models~\cite{Liu:2013if,Rostami:2013oq,PhysRevB.88.075409,:/content/aip/journal/adva/3/5/10.1063/1.4804936} and the $\bm k \cdot\bm p$ perturbation theory~\cite{Xiao:2012cr,Li:2013qf,Kormanyos:2013dq,PhysRevX.4.011034}. For our purposes it is enough to use the simplest possible, two-band (or four-band, if spin is included) $\bm k \cdot \bm p$ method~\cite{Xiao:2012cr}. To establish the form of effective $\bm k \cdot \bm p$ Hamiltonian we make use of symmetry arguments and the method of invariants.

\begin{figure}[t]%
\includegraphics*[width=\linewidth]{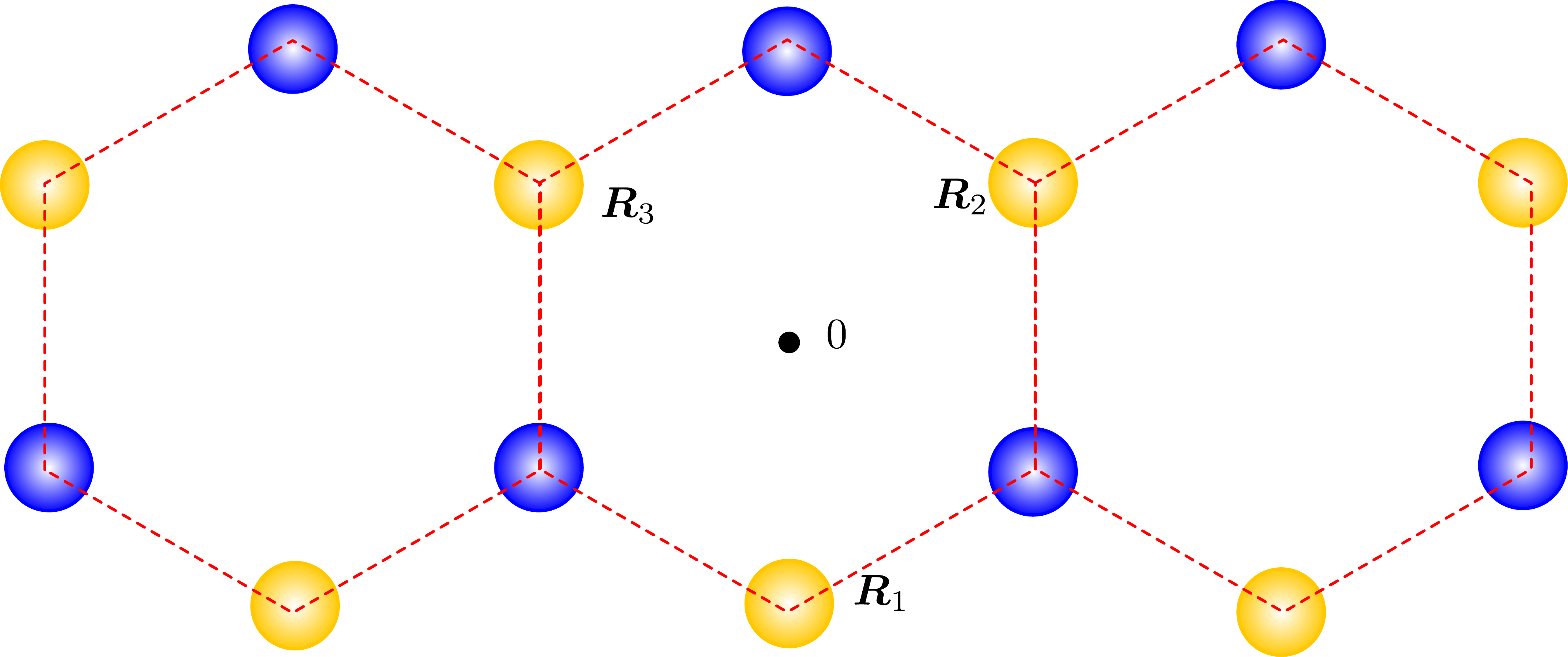}
\caption{%
 Schematic illustration of atoms in hexagonal lattice of MX$_2$. Yellow circles denote positions of transition metal M, blue circles show the projections on the monolayer plane of the chalcogens X; $0$ denotes the origin of point transformations, $\bm R_1$, $\bm R_2$ and $\bm R_3$ are the position vectors of metal atoms.}
\label{fig:atoms}
\end{figure}

Figure~\ref{fig:atoms} represents the positions of atoms in the hexagonal lattice of MX$_2$ projected onto the monolayer plane. The symmetry of the monolayer is described by the point group {D}$_{3h}$ which includes the horizontal reflection plane containing metals, $\sigma_h$, and three-fold rotation axis $C_3$ intersecting the horizontal plane in the center of hexagon. The complete list of {D}$_{3h}$ elements includes also the mirror-rotation axis $S_3$, three two-fold rotation axes $C_2$ lying in the monolayer plane and mirror reflection planes $\sigma_v$ which contain the $C_2$ axes. 

\begin{figure}[t]%
\includegraphics*[width=\linewidth]{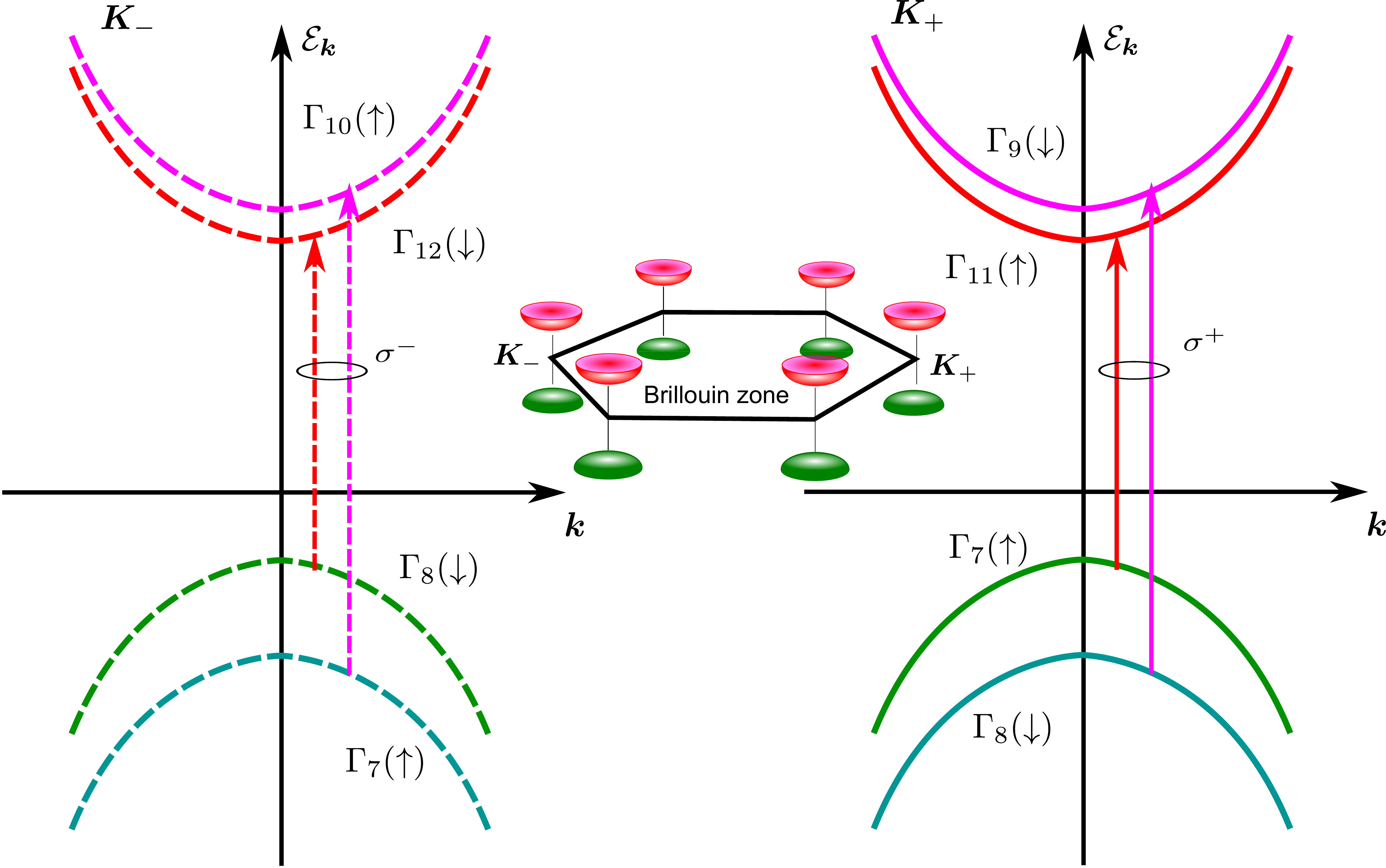}
\caption{%
 Sketch of the band structure of transition metal dichalcogenide monolayers. The bands are labelled by the corresponding irreducible spinor representations with arrows in parentheses indicating the electron spin orientation. Solid and dashed arrows show the transitions active under the normal incidence in the $\sigma^+$ and $\sigma^-$ polarizations, respectively. Inset shows the Brillouin zone and the band dispersion. The order of conduction band states is shown for MoS$_2$ in accordance with Ref.~\cite{Liu:2013if}.}
\label{fig:bands}
\end{figure}

The Brillouin zone of MX$_2$ monolayers is also hexagonal and the direct band gaps are realized at the points $\bm K_\pm$ corresponding to its edges, see inset in Fig.~\ref{fig:bands}. These two valleys are coupled by the time-reversal symmetry. The wave vector group at the $\bm K_\pm$ points is C$_{3h}$. The symmetry of conduction and valence band wavefunctions can be established taking into account that the conduction band is formed, neglecting spin and spin-orbit coupling, mostly from the $d_{z^2}$ orbitals of metal, while the valence band is formed mostly from the $d_{(x\pm \mathrm i y)^2}$ orbitals~\cite{Xiao:2012cr,PhysRevB.87.155304}. Here $x$ and $y$ are the Cartesian axes in the monolayer plane and $z$ is the sample normal. The state symmetry at the $\Gamma$ point is completely determined by that of the atomic orbitals. By contrast, the symmetry behavior of the states at the edges $\bm K_{\pm}$ of the Brillouin zone is determined not only by the symmetry of the particular orbitals but also by the relation between the phase ${\bm K_\pm}\cdot {\bm R}_j$ of the Bloch function and the phase $\left(\hat C_3 {\bm K_\pm}\right)\cdot {\bm R}_j$, where ${\bm R}_j$ is the orbital position and $\hat C_3$ is the three-fold rotation operator. As a result~\cite{Kormanyos:2013dq}, the valence band orbital Bloch functions being $\propto d_{(x\pm \mathrm i y)^2}$ in the $\bm K_+$ and $\bm K_-$ valleys, respectively, transforms according to the invariant representation of the C$_{3h}$ point group, $A'$ in notations of Ref.~\cite{Kormanyos:2013dq} or $\Gamma_1$ in the notations of Ref.~\cite{koster63}. The conduction band states transform in the valleys $\bm K_+$ and $\bm K_-$ according to $E_1'$ ($\Gamma_2$) like $x+\mathrm i y$, and $E_2'$ ($\Gamma_3$), like $x-\mathrm i y$, respectively. This symmetry analysis demonstrates that selection rules for the interband optical transition at the normal incidence of radiation are \emph{chiral}: The transitions in the $\bm K_+$ valley are active in the $\sigma^+$ polarization of light, while the transitions at $\bm K_-$ are excited by the $\sigma^-$ polarized light~\cite{Cao:2012fk,Mak:2012qf,Zeng:2012ys}.

Inclusion of the electron spin and spin-orbit coupling does not change these chiral selection rules. Indeed, taking into account the electron spin, the valence band states transform according to the spinor representations $\Gamma_7$, spin-up, and $\Gamma_8$, spin-down, in each valley, see Fig.~\ref{fig:bands}. The conduction band states in the $\bm K_+$ valley transform according to $\Gamma_{11}$, spin-up, and $\Gamma_9$, spin-down, while the $\bm K_-$ valley conduction band states transform according to $\Gamma_{10}$, spin-up, and $\Gamma_{12}$, spin-down, respectively. These spinor representations of the C$_{3h}$ point group are non-degenerate~\cite{koster63} which means that, in the absence of external forces, both the conduction and valence bands already are spin split at the $\bm K_{\pm}$ points. The time-reversal symmetry ensures the splittings to be of opposite signs in the $\bm K_+$ and $\bm K_-$ valleys. Therefore, as shown in Fig.~\ref{fig:bands}, the order of spin states at the $\bm K_+$ and $\bm K_-$ edges of the Brillouin zone is reversed. The spin splittings in each valley are sizeable: For the valence band states their order of magnitude is $\sim 100$~meV increasing with the atomic number of constitutive elements, while for the conduction band the splittings are in the meV to tens of meV range~\cite{Xiao:2012cr,Kormanyos:2013dq,Liu:2013if,PhysRevB.88.245436,PhysRevX.4.011034}. The dipole interaction with electromagnetic field  conserves the spin, therefore, there are only two allowed interband optical transitions under the normal incidence at each valley, as illustrated in Fig.~\ref{fig:bands} by vertical arrows. The selection rules are determined by the orbital part of the Bloch functions. Hence, the transitions in the $\bm K_\pm$ valleys are activated by $\sigma^\pm$-polarized photons, respectively.

In what follows we focus on the optical transitions involving the bottom conduction and topmost valence bands, i.e., the states $\Gamma_7$ and $\Gamma_{11}$ in the $\bm K_+$ valley and the states $\Gamma_8$ and $\Gamma_{12}$ in the $\bm K_-$ valley. Therefore, it is enough to resort to the two-band model where the effective $\bm k\cdot \bm p$ Hamiltonians in the vicinity of $\bm K_\pm$ points can be presented as $2\times 2$ matrices, respectively, 
\begin{equation}
\label{Hkp}
\mathcal H_+ =\begin{pmatrix}
E_g & \gamma_3 k_- \\
\gamma_3 k_+ & 0 
\end{pmatrix}, \quad \mathcal H_- =\begin{pmatrix}
E_g & {-}\gamma_3 k_+ \\
{-}\gamma_3 k_- & 0
\end{pmatrix}.
\end{equation}
Here the energy is referred to the valence band top, $E_g$ is the band gap, $k_\pm = k_x \pm\mathrm i k_y$, $\bm k = (k_x, k_y)$ is a two-dimensional wave vector of the electron reckoned from the corresponding Brillouin zone edge, and the real constant $\gamma_3$ is related with the interband matrix element of the electron momentum. Note that the choice of signs in the off-diagonal terms in $\mathcal H_\pm$ as well as the fact that $\gamma_3$ parameter is real is related with a definite convention about the phases of the Bloch functions and action of the time-reversal operator. In the two-band approximation neglecting contributions from distant bands and from bare electron dispersion~\cite{2015arXiv150304105W}  the conduction and valence bands are symmetric with the effective masses $m_c = -m_v \equiv m^*$, where
\begin{equation}
\frac{1}{m^*} = \frac{2\gamma_3^2}{\hbar^2E_g}.
\end{equation}
In MX$_2$ monolayers the effective mass $m^*$ is about $0.5m_0$, where $m_0$ is the free electron mass. More elaborate $\bm k \cdot \bm p$ models are presented in Refs.~\cite{Xiao:2012cr,Kormanyos:2013dq,PhysRevX.4.011034,Kormanyos:2014b,2015arXiv150304105W} and include remote bands into the consideration leading to the electron-hole  asymmetry and a better agreement with atomistic calculations. For instance, Ref.~\cite{Kormanyos:2014b} reports $m_c=0.56 m_0$, $m_v=-0.59 m_0$ for MoSe$_2$ and $m_c=0.28 m_0$, $m_v=-0.36 m_0$ for  WSe$_2$. It shows that the above two-band approximation can be used if a moderate accuracy for effective masses is needed. Overall, existing $\bm k \cdot \bm p$ models~\cite{Kormanyos:2014b} provide adequate results for the dispersion of electrons and holes in the vicinity of $\bm K_\pm$ points, but do not reproduce the observed Zeeman splitting for excitons and trions~\cite{2015arXiv150304105W}. However, the fine structure of the radiative doublet of the excitonic states, which is the focus of this paper, is not affected by the details of electron and hole bands. Hence, hereafter we use the simplest possible band structure models that will be particularly useful to describe the valley polarization dynamics.

\section{Exciton states in MX$_2$ monolayers}\label{sec:exc}

Relatively heavy masses of charge carriers and weak screening in two-dimensional systems yield strong excitonic effects, see Ref.~\cite{Yu30122014} for recent review. It is worth mentioning, that strong Coulomb effects in MX$_2$ monolayers make it possible to observe, besides neutral excitons, other Coulomb correlated complexes such as trions (also known as charged excitons)~\cite{Mak:2013lh} and biexcitons~\cite{biexciton}. The spin and valley dynamics of these complexes deserves further experimental and theoretical study and is not addressed here.

In the following we assume that the main contribution to the exciton wavefunctions is provided by the conduction and valence band states in the vicinity of $\bm K_\pm$ points of the $\bm k$-space and denote the conduction (valence) band branches by the spin index $s_e$ ($s_h$) and the valley index $\tau_e$ ($\tau_h$). 
Then the exciton state in a transition metal dichalcogenide monolayer can be presented within the $\bm k \cdot \bm p$ formalism as~\cite{ivchenko05a}
\begin{equation}
\label{exciton:kp}
\sum\limits_{\bm k_e,\bm k_h} C_{n;s_e \tau_e;s_h \tau_h}(\bm k_e,\bm k_h) |s_e\tau_e\bm k_e;s_h\tau_h\bm k_h\rangle,
\end{equation}
where $n=1s, 2p, \ldots$ denotes the states of electron-hole relative motion, $\bm k_e$ ($\bm k_h$) is the electron (hole) wavevector, and $|s_e\tau_e\bm k_e;s_h\tau_h\bm k_h\rangle$ denotes the excited state of the MX$_2$ monolayer with the occupied conduction band state $|s_e\tau_e\bm k_e\rangle$ and the empty valence band state $\hat{\mathcal K} |s_h\tau_h\bm k_h\rangle$, with $\hat{\mathcal K}$ being the time-reversal operator. It is convenient to present this two-particle excited state in the coordinate representation as a product~\cite{ivchenko05a,birpikus_eng}
\begin{equation}
\label{prod}
|s_e\tau_e\bm k_e;s_h\tau_h\bm k_h\rangle = U_{s_e\bm k_e}^{\tau_e}(\bm r_e) U_{s_h\bm k_h}^{(h)\tau_h}(\bm r_h),
\end{equation} 
where $U_{s_e\bm k_e}^{\tau_e}(\bm r_e)$ and $U_{s_h\bm k_h}^{(h)\tau_h}(\bm r_h)$ are the conduction and valence band Bloch functions, the latter taken in the hole representation. As functions of $\bm k_e$ and $\bm k_h$, the coefficients $C_{n;s_e\tau_e;s_h\tau_h}(\bm k_e,\bm k_h)$ are the Fourier transforms of exciton real-space envelope functions $C_{n;s_e\tau_e;s_h\tau_h}(\bm r_e,\bm r_h)$. 

Neglecting the band mixing the exciton states can be labelled by the center of mass wavevector $\mathbf K=(K_x,K_y)$, the envelope index $n$ as well as by the spin and valley indices of electrons and holes. The substantial spin splitting of the valence band allows us to consider independently A- and B-excitons formed of holes, respectively, in the top/bottom spin valence subbands, see Fig.~\ref{fig:bands}. In what follows we focus on the A-exciton states.

\begin{figure}[t]%
\includegraphics*[width=\linewidth]{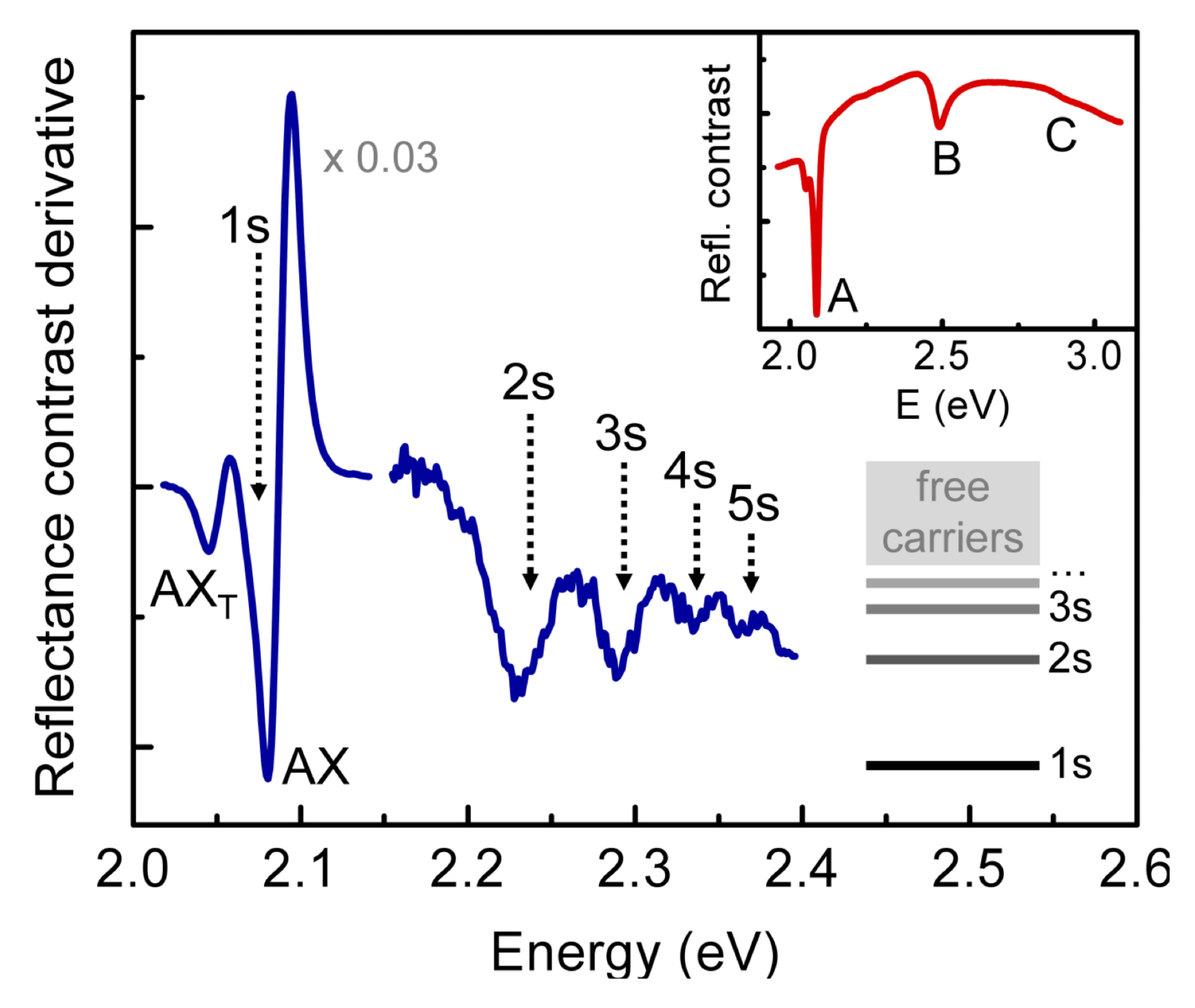}
\caption{The derivative of the reflectance contrast spectrum, $d(\Delta R/R)/d\omega$, of the WS$_2$ monolayer. The exciton ground state and the higher excited states are labeled by their respective quantum numbers (schematically shown at bottom-right corner). The spectral region around the $1s$ transition (AX) and the trion peak (AX$_{\rm T}$) of the A exciton is scaled by a factor of 0.03 for clarity. Inset shows the as-measured reflectance contrast $\Delta R/R$, allowing for the identification of the A, B and C transitions. From Ref.~\cite{Chernikov:2014a}.}
\label{fig:exc:refl}
\end{figure} 

Estimates in the framework of Wannier-Mott model of two-dimensional exciton give binding energies  $E_b = 2\mu e^4/(\varkappa\hbar)^2\sim 500$~meV for the reduced mass $\mu = m^*/2=0.25m_0$ and the background dielectric constant $\varkappa=5$. The latter follows from the assumption that the two-dimensional layer is deposited on a semi-infinite substrate with the static dielectric constant $10$.  An accurate evaluation of the exciton binding energy is a very complex problem, mainly due to non-trivial screening effects and the necessity to account for  fine details of the band structure. Various theoretical approaches have already been developed see, e.g.,~\cite{Ramasubramaniam:2012qa,PhysRevB.86.241201,PhysRevB.87.155304,Qiu:2013fe,Berkelbach:13,Berg:14}, which report values of $E_b$ in the range of $\sim 100$~meV to $\sim 1$~eV depending on the material, dielectric surrounding and level of approximations used. Recent works based on reflectivity measurements~\cite{Chernikov:2014a} and two-photon absorption~\cite{2014arXiv1404.0056W} indeed report the A-exciton binding energies in the range of above estimation for transition metal dichalcogenides monolayers. Figure~\ref{fig:exc:refl} shows the derivative of the reflectance contrast spectrum, $d(\Delta R/R)/d\omega$ of a WS$_2$ monolayer measured in Ref.~\cite{Chernikov:2014a}. Here $\Delta R$ is the difference of the reflection coefficients of the sample and substrate, $R$ is the substrate reflection coefficient, $\omega$ is the frequency of the light. The data clearly demonstrate $s$-shell excitonic series up to the principle quantum number $5$. The analysis carried out in Ref.~\cite{Chernikov:2014a} reveals strong deviations from the two-dimensional hydrogen-like model which is attributed to the complex screening of the Coulomb interaction in the MX$_2$ monolayers. The binding energy of exciton in WS$_2$ was found to be about $320$~meV in close agreement with theoretical estimates~\cite{Chernikov:2014a}. The higher value of exciton binding energy $E_b \sim 700$~meV for the same material was extracted in Ref.~\cite{ye:nature}. A value of $E_b\sim 600$~meV, was reported in~\cite{2014arXiv1404.0056W} for WSe$_2$ from identification of $2s$ and $2p$ excitonic states by means of single- and two-photon spectroscopy and DFT-GW band gap calculation, though a smaller value of $\sim 370$~meV have been also reported~\cite{PhysRevLett.113.026803}. A similar value of the binding energy, $E_b\gtrsim 570$~meV, was reported for excitons in MoS$_2$ from the photocurrent spectroscopy~\cite{Klots:2014aa}. The spread of reported binding energies can be related with use of different experimental and sample preparation methods as well as with unprecise knowledge of the single-particle band gap in MX$_2$ systems. Importantly, all experimental and theoretical results clearly show that the optical properties are governed by robust excitons.

Now we proceed to the fine structure of excitonic states which is the main goal of the paper. For a given relative motion quantum number $n$ ($n=1s$ in what follows) and center of mass wavevector $\mathbf K$ we have eight A-exciton states in total, four of them are intra-valley (or direct) excitons with the occupied state in the conduction band and the empty state in the valence band being in the same valley (note that in this case $\tau_e = -\tau_h$ since $\tau_h$ refers to the hole valley index) and the others four are inter-valley (indirect) excitons. The latter can manifest themselves in optical transitions if a third particle, a phonon or an impurity, is involved in the transition. Moreover, it follows from the selection rules illustrated in Fig.~\ref{fig:bands} that in the dipole approximation only two of four intra-valley excitons are coupled to the light under normal incidence: those are the states with $s_e = - s_h$ (again, spins of empty state in the valence band and in the conduction band are the same). The remaining two states are dark at the normal incidence. Note, that the A-excitons with $s_e=s_h$ (spin-forbidden) are, due to the spin-orbit mixing, active only in the $z$-polarization~\cite{glazov2014exciton}. In spite of being dark, these states can play an important role in the photoluminescence formation~\cite{zhang:APS}. Other dark states, such as $2p$ excitons are visible in two-photon absorption processes~\cite{2014arXiv1404.0056W,ye:nature}. These effects are beyond the scope of the present paper.

\begin{figure}[t]%
\includegraphics*[width=\linewidth]{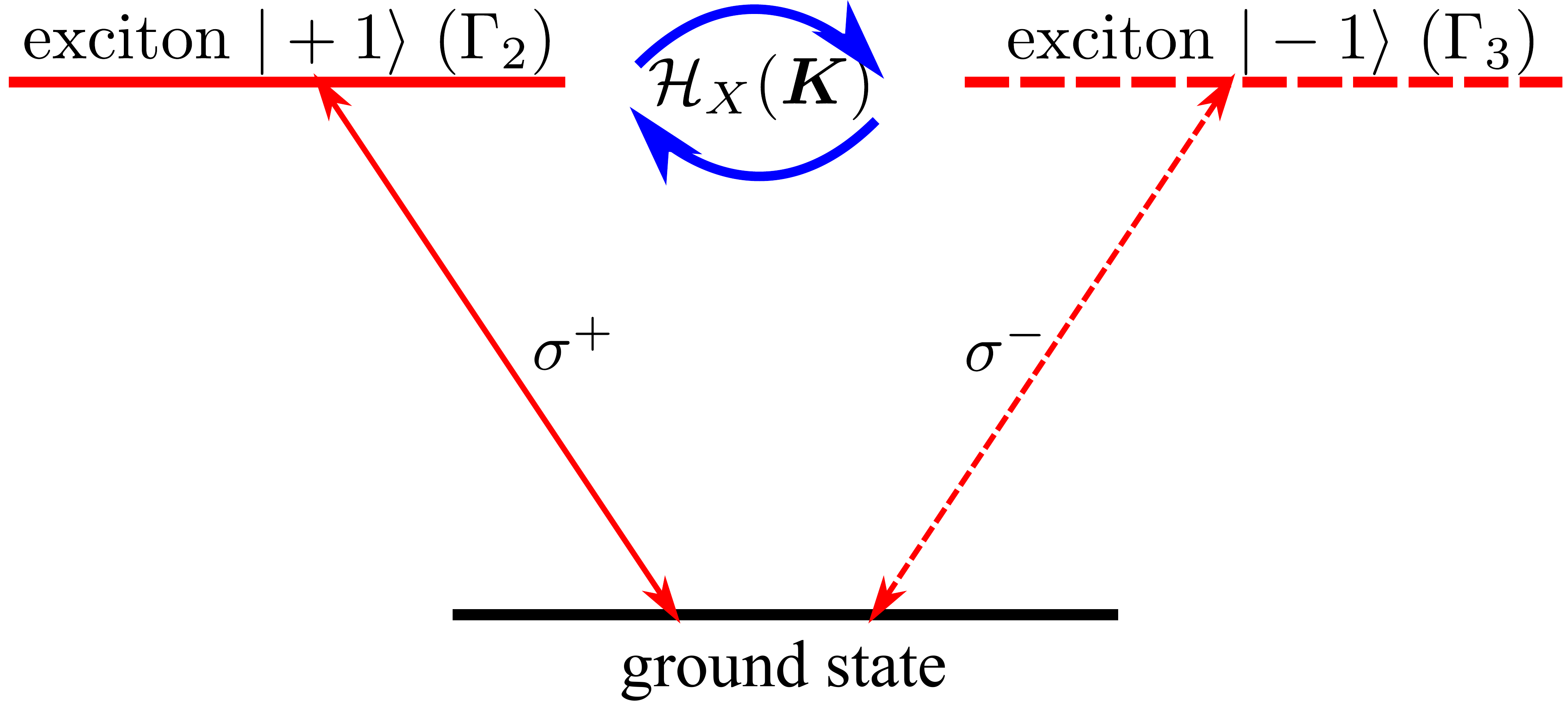}
\caption{Optical selection rules of the two bright A-exciton states with the small center of mass wavevector $\mathbf K$ and their long-range Coulomb exchange coupling, Eq.~\eqref{H:bright}.}
\label{fig:exc:states}
\end{figure}

\begin{figure}[t]%
\includegraphics*[width=0.99\linewidth]{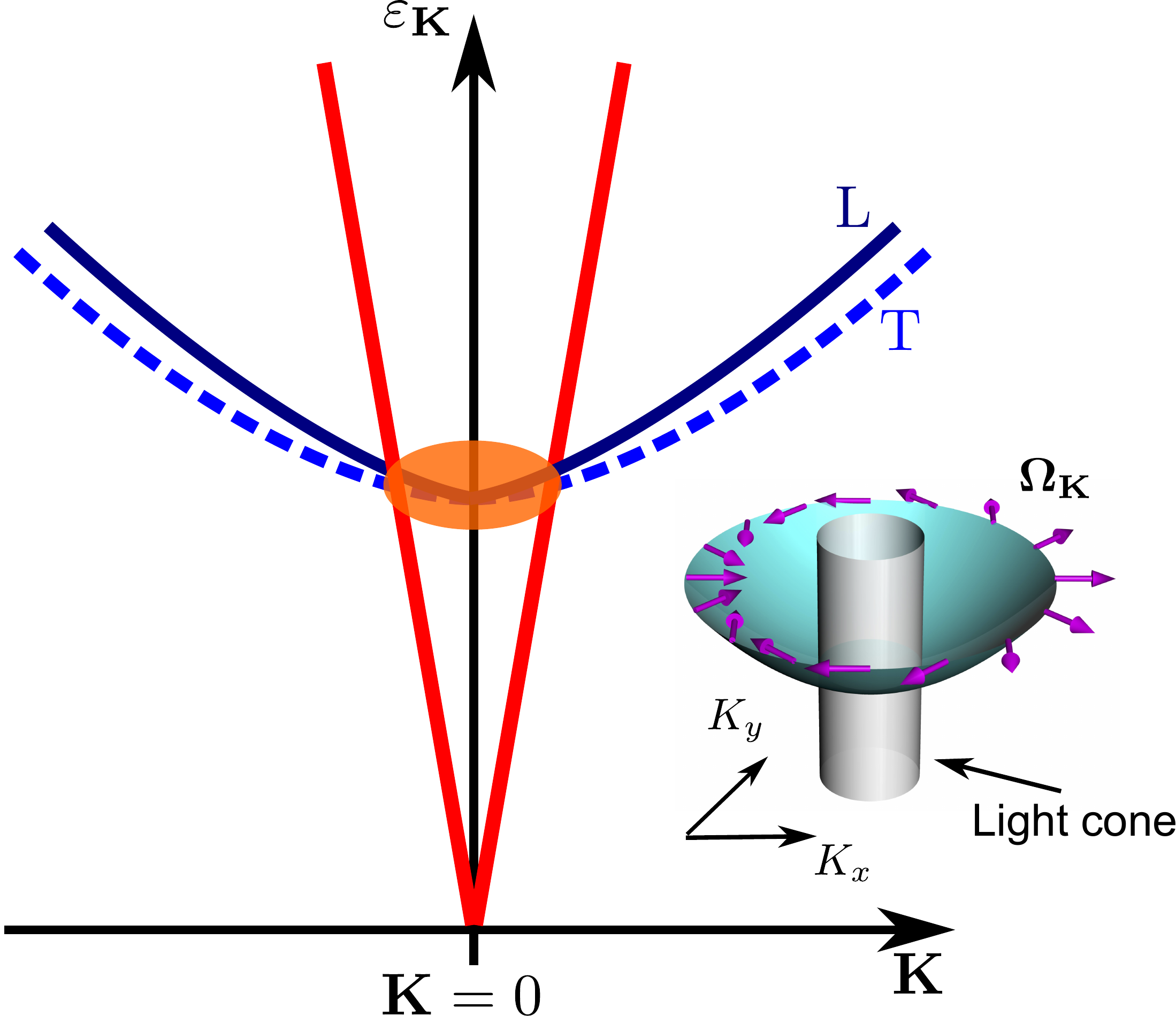}
\caption{Schematic illustration of exciton dispersion showing splitting of the bright doublet into linearly polarized eigenstates with the dipole moment oriented parallel ({L}) and perpendicular ({T}) to the in-plane wavevector. $\mathbf K=0$ point corresponds to center of exciton Brillouin zone. Red solid lines show photon dispersion (light cone). Dispersion is shown not to scale. Inset shows orientation of effective field $\bm \Omega_{\mathbf K}$ vs. wavevector $\mathbf K$ direction in Eq.~\eqref{H:bright}, the splitting is not shown.}
\label{fig:split}
\end{figure}

\section{Fine structure of bright excitons}\label{sec:fine}

Bright A-excitonic states in the $\bm K_+$ and $\bm K_-$ valleys transform according to the $\Gamma_2$ and $\Gamma_3$ representations of $\mathrm C_{3h}$ point symmetry group and are excited by $\sigma^+$ and $\sigma^-$ photons, respectively, see Fig.~\ref{fig:exc:states} for illustration. This pair of states can be also related with the two dimensional representation $\Gamma_6$ of the $\mathrm D_{3h}$ point group relevant for the overall symmetry of the MX$_2$ monolayer. The symmetry arguments show that for excitons propagating in the structure plane with the center of mass wave vector $\mathbf K = (K_x,K_y)$ the eigenstates are linear combinations of the circularly polarized states corresponding to the microscopic dipole moment of exciton oscillating parallel or perpendicular to its wave vector~\cite{maialle93,ivchenko97pss,goupalov98,ivchenko05a}. The effective Hamiltonian acting in the basis of $\Gamma_2$ and $\Gamma_3$ states can be written in the framework of invariants method as a $2\times 2$ matrix:
\begin{multline}
\label{H:bright}
\mathcal H_{X} (\mathbf K)=  
\begin{pmatrix}
0 & \alpha (K_x - \mathrm i K_y)^2 \\
 \alpha  (K_x+ \mathrm i K_y)^2 & 0
\end{pmatrix} \\
= \frac{\hbar}{2} \left( \bm \Omega_{\mathbf K} \cdot \bm \sigma \right).
\end{multline}
Here $\alpha \equiv \alpha(K)$ is the coupling parameter to be determined from the microscopic model. The second line in Eq.~\eqref{H:bright} gives the effective Hamiltonian in the pseudospin representation, with $\bm \sigma = (\sigma_x,\sigma_y)$ being the vector composed of  Pauli matrices and $\bm \Omega_{\mathbf K}$ is the  effective pseudospin precession frequency also termed as effective ``magnetic-like'' field acting on an exciton pseudospin $1/2$. It  has non-zero components $\hbar \Omega_{x} = 2\alpha K^2\cos{2\vartheta}$ and $\hbar \Omega_{y} = 2\alpha K^2 \sin{2\vartheta}$, where $\vartheta$ is the angle between $\mathbf K$ and the in-plane axis $x$. The effective field is described by second angular harmonics of the exciton wave vector $\mathbf K$, this is because a transfer of angular momentum $2$ is needed to flip exciton circular polarization. The dependence of the spin precession frequency $\bm \Omega_{\mathbf K}$ on the direction of the exciton center of mass wavevector $\mathbf K$ is schematically presented in the inset to Fig.~\ref{fig:split}. The splitting between the eigenmodes is $\hbar\Omega_{K} = 2\alpha K^2$; it corresponds to the exciton longitudinal transverse splitting and is depicted in Fig.~\ref{fig:split}.

The coupling parameter $\alpha(K)$ in Eq.~\eqref{H:bright} is determined, in fact, by the \emph{long-range exchange interaction} between an electron and a hole~\cite{birpikus_eng,denisovmakarov,ivchenko05a}. It can be found either in the framework of $\bm k\cdot \bm p$ perturbation theory or in the electrodynamical approach. In the latter case the interaction of the mechanical exciton, i.e., the electron-hole pair bound by the direct Coulomb interaction~\cite{agr_ginz}, with the macroscopic longitudinal electric field should be taken into account. We start with the electrodynamical treatment and then demonstrate its equivalence to the $\bm k\cdot \bm p$ calculation.

In the first approach the exciton frequencies can be found from the poles of the reflection coefficient of the two-dimensional structure~\cite{agr_ginz}. Let us derive the reflection coefficient of the MX$_2$ monolayer at the frequency range close to the lowest in energy excitonic transitions (A-excitons). For simplicity we consider the monolayer situated in $(xy)$ plane surrounded by dielectric media with high-frequency a dielectric constant $\varkappa_b$, the contrast of background dielectric constants between the material and surroundings is disregarded. The allowance for the dielectric constant as well as for the substrate is straightforward following Ref.~\cite{ivchenko05a}, see below. The geometry is illustrated in Fig.~\ref{fig:geom}.

\begin{figure}[b]
\includegraphics[width=0.75\linewidth]{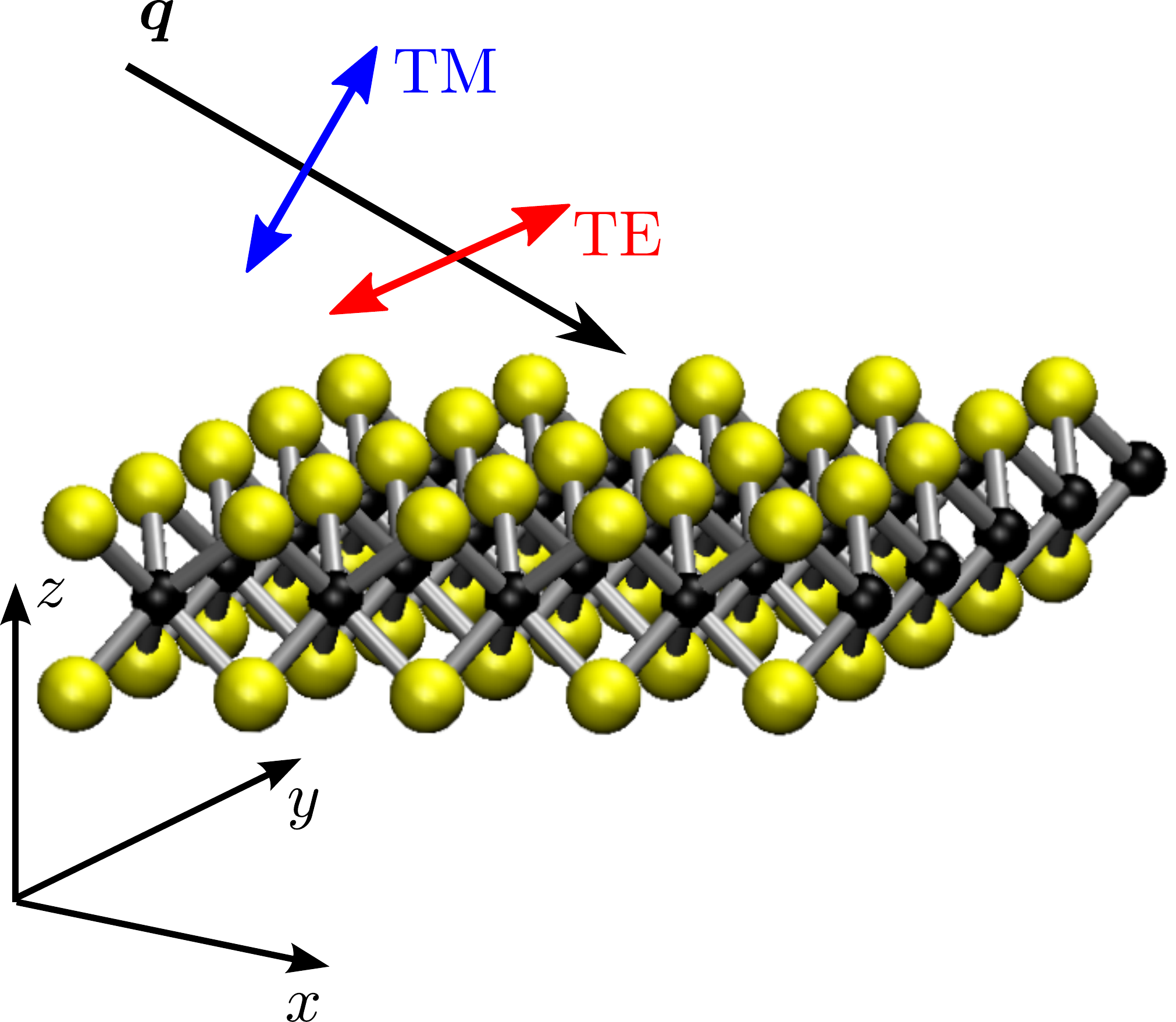}
\caption{Schematic illustration the system geometry with $s$ (TE mode) and $p$ (TM mode) polarized incident light.}\label{fig:geom}
\end{figure}

Maxwell equations for the electric field vector $\bm E$ can be recast as~\cite{ivchenko05a}
\begin{equation}
\label{maxwell0}
\rot\rot \bm E = \left(\frac{\omega}{c}\right)^2\left[\varkappa_b\bm E + 4\pi \bm P(z)\right],
\end{equation}
where $\omega$ is the frequency of light and $\bm P(z)$ is the excitonic contribution to the dielectric polarization. 
We solve Maxwell equations~\eqref{maxwell0} the for electromagnetic field taking into account the excitonic contribution to the dielectric polarization. The latter is calculated within the first-order perturbation theory using dipole approximation for exciton-light coupling and assuming Wannier-Mott exciton as~\cite{ivchenko05a,glazov2014exciton}:
\begin{equation}
\label{nonloc}
\bm P(z) = \frac{ \delta(z) |\varphi(0)|^2 \bm E(z)}{\omega_0 - \omega + \mathrm i \Gamma} \frac{e^{2}\gamma_3^2}{\hbar^3\omega_0^2}.
\end{equation} 
Here $\omega$ is the incident radiation frequency, $\omega_0$ is the exciton resonance frequency determined by the band gap and its binding energy, $\Gamma$ is its nonradiative damping, $\varphi(\rho)$ is the relative electron-hole motion wavefunction in the exciton, and $\gamma_3$ is defined in Eq.~\eqref{Hkp}. 
The short-range part of the exchange interaction contribution can also be included in $\omega_0$. Factor $\delta(z)$ ensures that the dipole moment is induced only in the monolayer of transition metal dichalcogenide whose width is negligible compared with the light wavelength. At a normal incidence of radiation the amplitude reflection coefficient of a monolayer has a standard form~\cite{andreani91,ivchenko05a} 
\begin{equation}
\label{normal}
r(\omega) = \frac{\mathrm i \Gamma_0}{\omega_0 - \omega - \mathrm i (\Gamma_0+ \Gamma)},
\end{equation}
where 
\begin{equation}
\label{Gamma0}
\Gamma_0 = \frac{2\pi q e^2 \gamma_3^2}{\hbar^3\varkappa_b \omega_0^2} {|\varphi(0)|^2}, 
\end{equation}
is the radiative decay rate of an exciton in the monolayer, $q= \sqrt{\varkappa_b} \omega/c$ is the wavevector of radiation. Rough estimate of radiative lifetime for the Wannier-Mott exciton parameters in MX$_2$ gives $1/(2\Gamma_0)\sim 1$~ps in agreement with recent measurements~\cite{PhysRevLett.112.047401,KornMoS2,PhysRevB.90.075413}. Note, that $\Gamma_0$ in transition metal dichalcogenides exceeds by far the radiative decay rate for excitons in conventional semiconductor quantum wells~\cite{Vinattiery,maialle93,Dareys1993353,vina99}
The parameters of the pole in the reflectivity, Eq.~\eqref{normal}, describe the eigenenergy and decay rate of the exciton with allowance for the light-matter interaction.
We note that in the approximation of strictly two-dimensional layer the mechanical exciton eigenfrequency is not renormalized. In this approximation the  transmission coefficient amplitude is equal to $t(\omega) = 1+r(\omega)$. In agreement with overall $\mathrm D_{3h}$ symmetry of the MX$_2$ monolayer, the reflection coefficient at a normal incidence is polarization independent. 

In order to find the fine structure of the moving exciton energy spectrum we consider the  oblique incidence of radiation in the $(xz)$-plane, Fig.~\ref{fig:geom}. The solution of Maxwell equations~\eqref{maxwell0} yields two eigenmodes of electromagnetic field: $s$ (or TE) polarized wave with $\bm E\parallel y$ and perpendicular to the light incidence plane, and $p$ (TM) polarized  wave with $\bm E$ in the incidence plane, see Fig.~\ref{fig:geom}. The reflection coefficients in a given polarization $i=s$ or $p$ have, like Eq.~\eqref{normal}, the pole contributions with the modified parameters
\begin{equation}
\label{r:alpha}
r_i(\omega) = \frac{\mathrm i \Gamma_{0i}}{\omega_{0i} - \omega - \mathrm i (\Gamma_{0i}+ \Gamma)}, 
\end{equation}
where
\begin{equation}
\label{Gamma0s}
\Gamma_{0s} = \frac{q}{q_z} \Gamma_0, \quad \Gamma_{0p} = \frac{q_z}{q} \Gamma_0,
\end{equation}
$q_z = (q^2 - q_\parallel^2)^{1/2}$ is the $z$ component of the light wavevector, $q_\parallel$ is its in-plane component, and $\omega_{0i} \equiv \omega_0(q_\parallel)= \omega_0 + \hbar q_\parallel^2/(2M)$ is the mechanical exciton frequency, $M$ is the exciton effective mass.

The light-matter interaction results in the radiative decay of the excitons with the wavevectors inside the light cone, $q_\parallel \leqslant \sqrt{\varkappa_b} \omega/c$, see Fig.~\ref{fig:split}. The decay rates are different for {T}- and {L}-polarized excitons. For the excitons outside the light cone, $q_z$ becomes imaginary and corresponding exciton-induced electromagnetic field decays exponentially with the distance from the monolayer. Therefore, exciton interaction with the field results in renormalization of its frequency rather than its decay rate~\cite{goupalov98,ivchenko05a,goupalov:electrodyn}. Formally, it corresponds to imaginary $\Gamma_{0i}$ in Eqs.~\eqref{r:alpha}, \eqref{Gamma0s}. Making analytical continuation of Eqs.~\eqref{r:alpha} to the complex $q_z$, and 
introducing the notation $\mathbf K = \bm q_{\parallel}$ for the center of mass wavevector of an exciton, we obtain from the poles of reflection coefficients the splitting between the longitudinal ($\bm P \parallel \mathbf K$) and transverse ($\bm P \perp \mathbf K$) exciton states:
\begin{equation}
\label{LTsplitting}
\Delta E = \hbar \Gamma_0 \frac{K^2}{q \sqrt{K^2 - q^2}} \approx \hbar \Gamma_0 \frac{K}{q},
\end{equation}
where the approximate equation holds for $K\gg q$ and one can replace $\omega$ by $\omega_0$ in the definition of $q$ since exciton dispersion is almost flat as compared with the dispersion of light. Equation~\eqref{LTsplitting} for excitons outside the light cone is in agreement with the phenomenological Hamiltonian~\eqref{H:bright}. The comparison of the splittings enables us to determine the constant $\alpha$ in Eq.~\eqref{H:bright} as
\begin{equation}
\label{alpha}
\alpha(K)  = \frac{\hbar \Gamma_0}{2 K q}.
\end{equation}
The long-range exchange splitting of excitonic states scales approximately $\propto K$ for $K \gg q$, i.e. for the states outside the light cone. We stress that the states within the light cone undergo renormalization of damping, i.e. imaginary part of the energy, rather than of the renormalization of its real part.

Equation~\eqref{alpha} can be easily generalized for the realistic case of the system ``vacuum -- monolayer of MX$_2$ -- substrate'' under assumption of  the same background refractive index $n$ of the monolayer and the substrate. The reflection at the boundary ``vacuum -- monolayer'' yields the replacement in denominators of Eq.~\eqref{r:alpha} $\Gamma_{0,i} \to \Gamma_{0,i}(1+\tilde r_i)$, where background reflectivities $\tilde r_i$ in $s$ and $p$ polarizations are given by the Fresnel formula with the refractive index $n$~\cite{ll8_eng}. In this case in accordance with Ref.~\cite{PhysRevB.90.161302} 
\begin{equation}
\label{alpha1}
\alpha(K) =  \frac{c\hbar\Gamma_0  }{2K \omega_0} \frac{n+1}{n^2+1}.
\end{equation} 
Thus, the long-range exchange interaction constant $\alpha$ is determined by the radiative decay rate of the exciton and the geometry of the system. Therefore, is expected to be enhanced in transition metal dichalcogenides monolayers compared with conventional semiconductor structures due to high oscillator strengths of excitons in MX$_2$.

\begin{figure*}[t]
\includegraphics[width=0.48\linewidth]{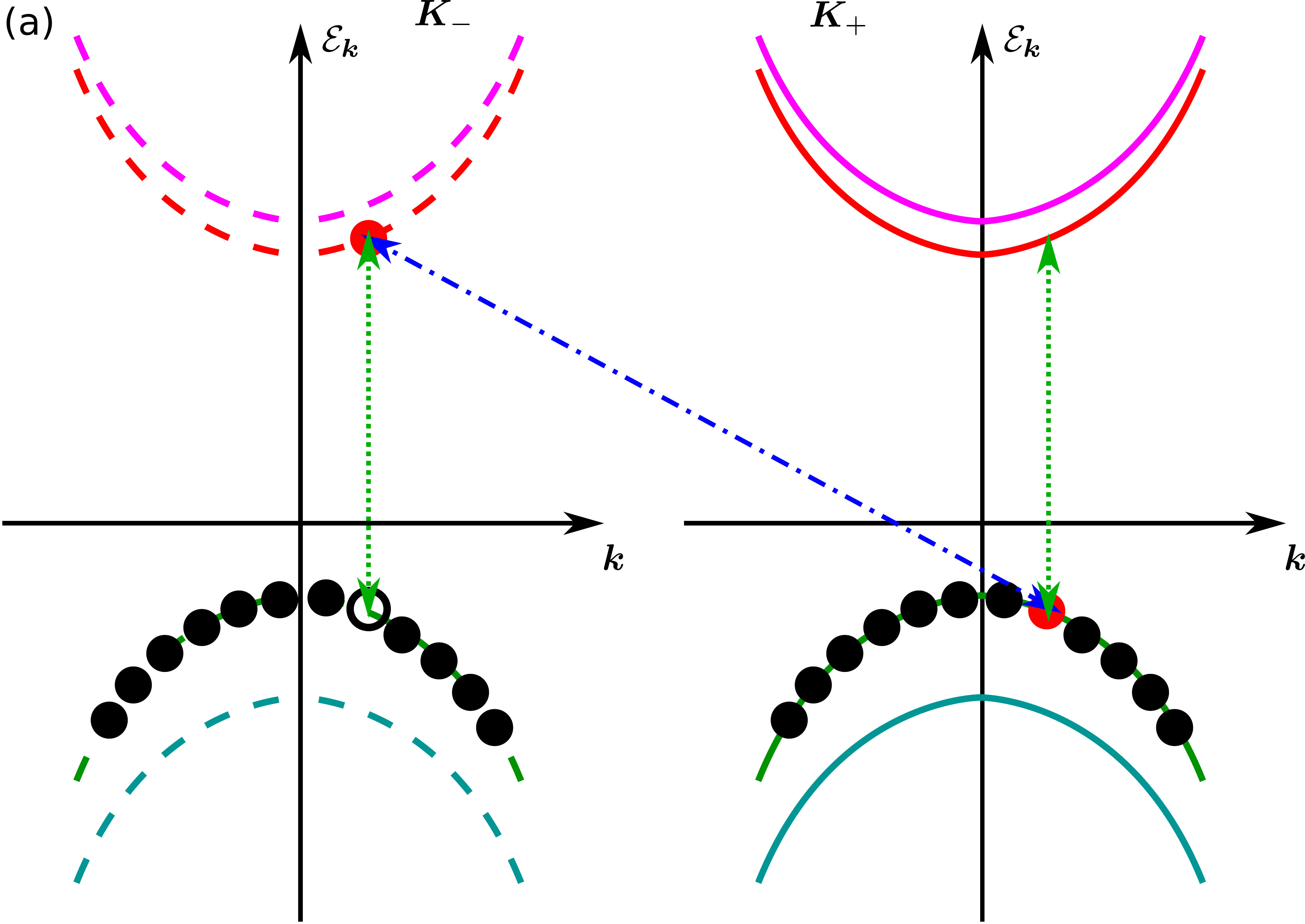}\hfill
\includegraphics[width=0.48\linewidth]{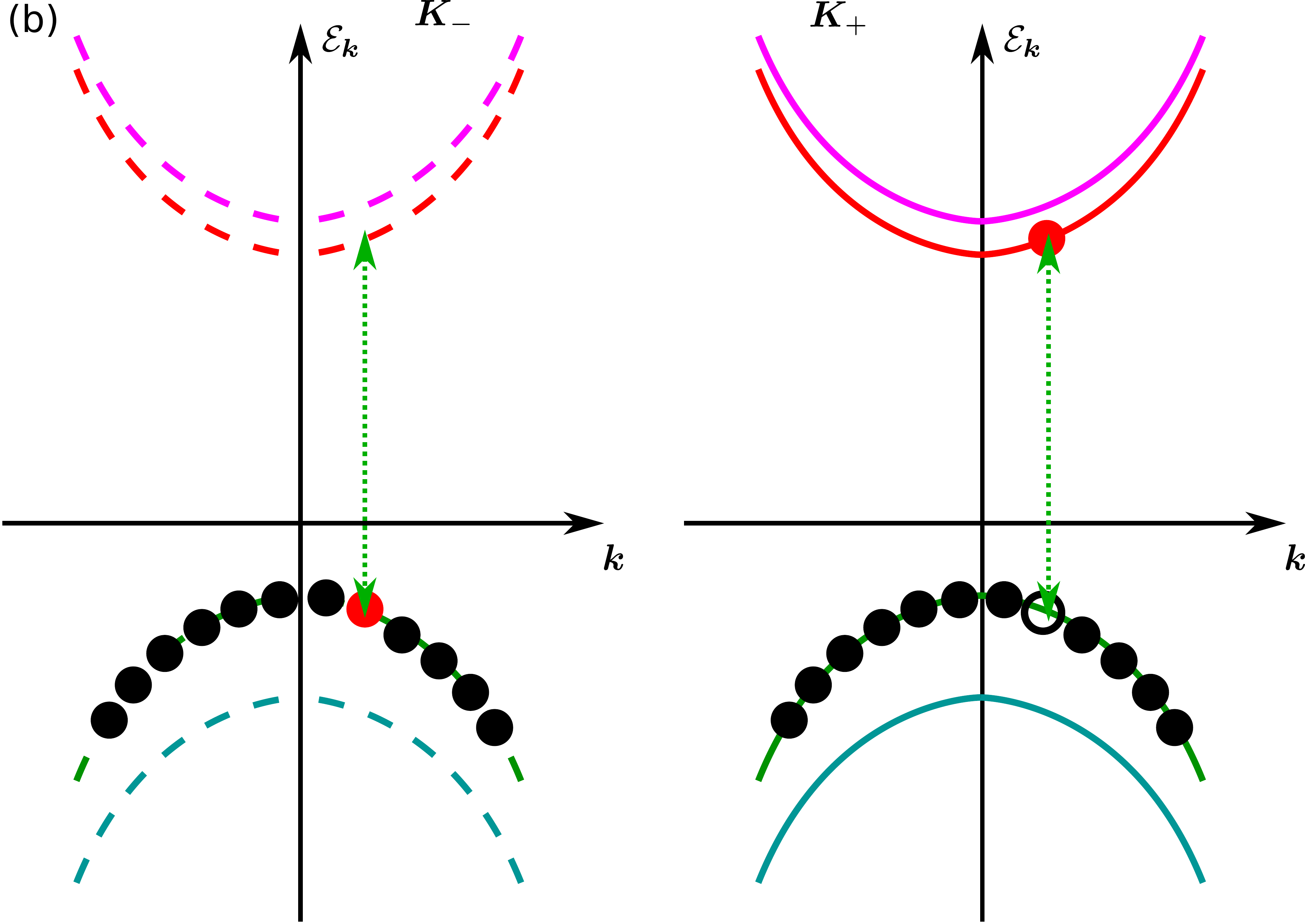}
\caption{Sketch of exciton intervalley transfer: (a) shows the initial state, (b) shows the final state. Filled circles denote electrons, empty circle denotes hole. Dash-dotted arrow shows the Coulomb interaction, green dotted lines show the $\bm k \cdot \bm p$ mixing.}\label{fig:Coul}
\end{figure*}

The electrodynamical treatment is general and is independent of the particular model of the band structure and of exciton wavefunction: The key parameter in Eqs.~\eqref{alpha}, \eqref{alpha1} is the radiative decay rate of exciton, $\Gamma_0$. It can be considered as a phenomenological parameter of the theory to be found from independent experiments. However, it is instructive to derive Eqs.~\eqref{H:bright} and \eqref{alpha} making use of $\bm k \cdot \bm p$ perturbation theory as well. It allows us to  illustrate how the Coulomb interaction enables the electron-hole pair transfer between the $\bm K_+$ and $\bm K_-$ valleys. As a first step, following Ref.~\cite{glazov2014exciton} we consider the exchange interaction between two electrons $\psi_m$ and $\psi_n$ occupying $\Gamma_{12}$ band in the $\bm K_-$ valley and $\Gamma_7$ band in the $\bm K_+$ valley with the wave vectors $\bm k_1$ and $\bm k_2$ reckoned from the edges of Brillouin zone, see red filled circles in Fig.~\ref{fig:Coul}(a). Making use of the Hamiltonians \eqref{Hkp} and taking into account $\bm k\cdot \bm p$ interaction in the first order we present their wavefunctions in the form
\begin{subequations}
\label{initial}
\begin{eqnarray}
\psi_m(\bm r_1) = \mathrm e^{\mathrm i \bm k_1 \bm r_1} \left[U_{12}^-(\bm r_1) {-} \frac{\gamma_3 k_{1,-}}{E_g} U_{8}^-(\bm r_1)\right],\\
\psi_n(\bm r_2) = \mathrm e^{\mathrm i \bm k_2 \bm r_2} \left[U_{7}^+(\bm r_2) - \frac{\gamma_3 k_{2,-}}{E_g} U_{11}^+(\bm r_2)\right].
\end{eqnarray}
\end{subequations}
Here $U_\Gamma^\tau(\bm r)$ are the Bloch amplitudes at the $\bm K_\tau$ points, $\tau=\pm$ enumerates valleys, subscript $\Gamma$ denotes the representation of $\mathrm C_{3h}$ point group corresponding to the particular Bloch function. As the final states of the electron pair we consider the conduction band $\Gamma_{11}$ state in the $\bm K_+$ valley, $\psi_{m'}$, and the valence band  $\Gamma_{8}$ state in the $\bm K_-$ valley, $\psi_{n'}$ characterized by the wavevectors $\bm k_1'$ and $\bm k_2'$, respectively. Similarly to Eqs.~\eqref{initial} their wavefunctions read, Fig.~\ref{fig:Coul}(b),
\begin{subequations}
\label{final}
\begin{eqnarray}
\psi_{m'}(\bm r_1) = \mathrm e^{\mathrm i \bm k_1' \bm r_1} \left[U_{11}^-(\bm r_1) {-} \frac{\gamma_3 k_{1,+}}{E_g} U_{7}^-(\bm r_1)\right],\\
\psi_{n'}(\bm r_2) = \mathrm e^{\mathrm i \bm k_2' \bm r_2} \left[U_{8}^+(\bm r_2) - \frac{\gamma_3 k_{2,+}}{E_g} U_{12}^+(\bm r_2)\right].
\end{eqnarray}
\end{subequations}
According to the general theory~\cite{birpikus_eng} the exchange matrix element of the Coulomb interaction $V(\bm r_1 - \bm r_2)$ can be written as
\begin{multline}
\label{exchange}
\langle m' n'| V(\bm r_1 - \bm r_2)| m n\rangle = -V_{\bm k_1' - \bm k_2} \delta_{\bm k_1 + \bm k_2, \bm k_1'+\bm k_2'} \\ \times {\frac{\gamma_3^2}{E_g^2}}(k_{2,-} - k_{2,-}')(k_{1,-}- k_{1,-}').
\end{multline}
Here $V_{\bm q}$ is the two-dimensional Fourier transform of the bare Coulomb potential. Equation~\eqref{exchange} clearly shows that the combined action of the $\bm k\cdot \bm p$ mixing and the Coulomb interaction enables intervalley transfer of the electron-hole excitation, see Fig.~\ref{fig:Coul}. As demonstrated in Fig.~\ref{fig:Coul} each electron remains in the same valley, but due to $\bm k \cdot \bm p$ mixing changes the band. Therefore, the long-range exchange interaction transfers the electron-hole pair between the $\bm K_-$ and $\bm K_+$ valleys of the energy spectrum.

As a second step of the derivation, in order to reproduce Eq.~\eqref{H:bright}, we make standard transformations from the electron-electron representation to the electron-hole representation in Eq.~\eqref{exchange}, see Eqs.~\eqref{exciton:kp}, \eqref{prod} and Ref.~\cite{birpikus_eng} and perform averaging over the relative motion wavefunction. Finally, inclusion of the high-frequency screening, see Refs.~\cite{goupalov98,zhilich72:eng,zhilich74:eng} for details, yields off-diagonal element $\langle \Gamma_2 |\mathcal H_X (\mathbf K)|\Gamma_3\rangle$ in Eq.~\eqref{H:bright}. The LT splitting in Eq.~\eqref{LTsplitting} in agreement with Eq.~\eqref{Gamma0} reads 
\begin{equation}
\label{exchange:bis}
\Delta E = \frac{2\pi e^2\hbar^2|\varphi(0)|^2}{\varkappa_b m_0^2\omega_0^2} \frac{(Kp_{cv})(Kp_{cv}^*)}{K}
\end{equation}
with $p_{cv}$ being interband momentum matrix element. We stress that the Coulomb interaction is long-range, it does not provide intervalley transfer of individual electrons, however, the exchange process involves one electron from the $\bm K_+$ and another one from the $\bm K_-$ valley. Equations \eqref{H:bright} and \eqref{LTsplitting} describe the long-range (also termed annihilation or resonant) exchange interaction between an electron and a hole. It was previously derived for excitons in semiconductor quantum wells both on the basis of electrodynamics and quantum mechanically, see Refs.~\cite{maialle93,goupalov98,ivchenko05a} and references therein. 

\section{Bright exciton spin dynamics. Theory}\label{sec:theor}

The dynamics of bright exciton doublet can be conveniently described within the pseudospin formalism~\cite{ivchenko05a}.  In this method, the $2\times 2$ exciton spin-density matrix $\varrho_{\mathbf K}$ in the basis of $\Gamma_2$ and $\Gamma_3$ states is decomposed as
\begin{equation}
\label{dens}
\varrho_{\mathbf K} = n_{\mathbf K} + \bm S_{\mathbf K} \cdot \bm \sigma,
\end{equation}
 with $n_{\mathbf K} = {\rm Tr}\{\varrho_{\mathbf K}/2\}$ being the spin-averaged distribution function of excitons, $\bm S_{\mathbf K} = {\rm Tr}\{\varrho_{\mathbf K}\bm \sigma/2\}$ being the pseudospin. Here unit $2\times 2$ matrix is omitted and $\bm \sigma = (\sigma_x,\sigma_y,\sigma_z)$ is a vector of Pauli matrices. Pseudospin components describe orientation of the oscillating microscopic dipole moment of exciton, particularly, $S_z/n$ gives the degree of circular polarization or valley polarization, $S_x/n$ and $S_y/n$ give the degree of linear polarization in two axes frames rotated by $45^{\rm o}$ with respect to each other, i.e. valley coherences or exciton alignment.

The exciton pseudospin satisfies the kinetic equation~\cite{maialle93,maialle00}
\begin{equation}
\label{kin}
\frac{\partial \bm S_{\mathbf K}}{\partial t} + \bm S_{\mathbf K} \times \bm \Omega_{\mathbf K} = \bm Q\{ \bm S_{\mathbf K} \},
\end{equation}
where $\bm \Omega_{\mathbf K}$ is defined by Eq.~\eqref{H:bright} and $\bm Q\{ \bm S_{\mathbf K} \}$ is the collision integral describing the variation of spin distribution function in the process of scattering. Its general form is given by the balance of in- and out-scattering processes:
\begin{equation}
\label{Q}
\bm Q\{ \bm S_{\mathbf K} \} = \sum_{\mathbf K'} \left(W_{\mathbf K\mathbf K'} S_{\bm K'} - W_{\mathbf K'\mathbf K} S_{\mathbf K}\right),
\end{equation}
where the scattering rates $W_{\mathbf K\mathbf K'}$ are determined by the particular interactions, i.e. elastic scattering by impurities and other defects, exciton-phonon interaction, exciton-exciton and exciton-electron scattering. Equation~\eqref{kin} shows that the long-range exchange interaction induced effective field $\bm \Omega_{\mathbf K}$ acts as a driving force for exciton spin dynamics. 

Similarly to the case of free excitons in quantum wells~\cite{maialle93,ivchenko05a} different regimes of spin decoherence in MX$_2$ monolayers can be identified depending on the relation between the characteristic spin precession frequency and scattering rates~\cite{Henn98}. To be specific, we address the situation where the excitons are assumed to be thermalized and distributed according to the Boltzmann function $n(\varepsilon) \propto \exp{(-\varepsilon/k_B T)}$, where $\varepsilon \equiv \varepsilon_{\mathbf K}$ is exciton dispersion, $k_B T$ is the exciton temperature measured in the energy units, $k_B$ is the Boltzmann constant. 

\begin{figure}[t]
\includegraphics[width=\linewidth]{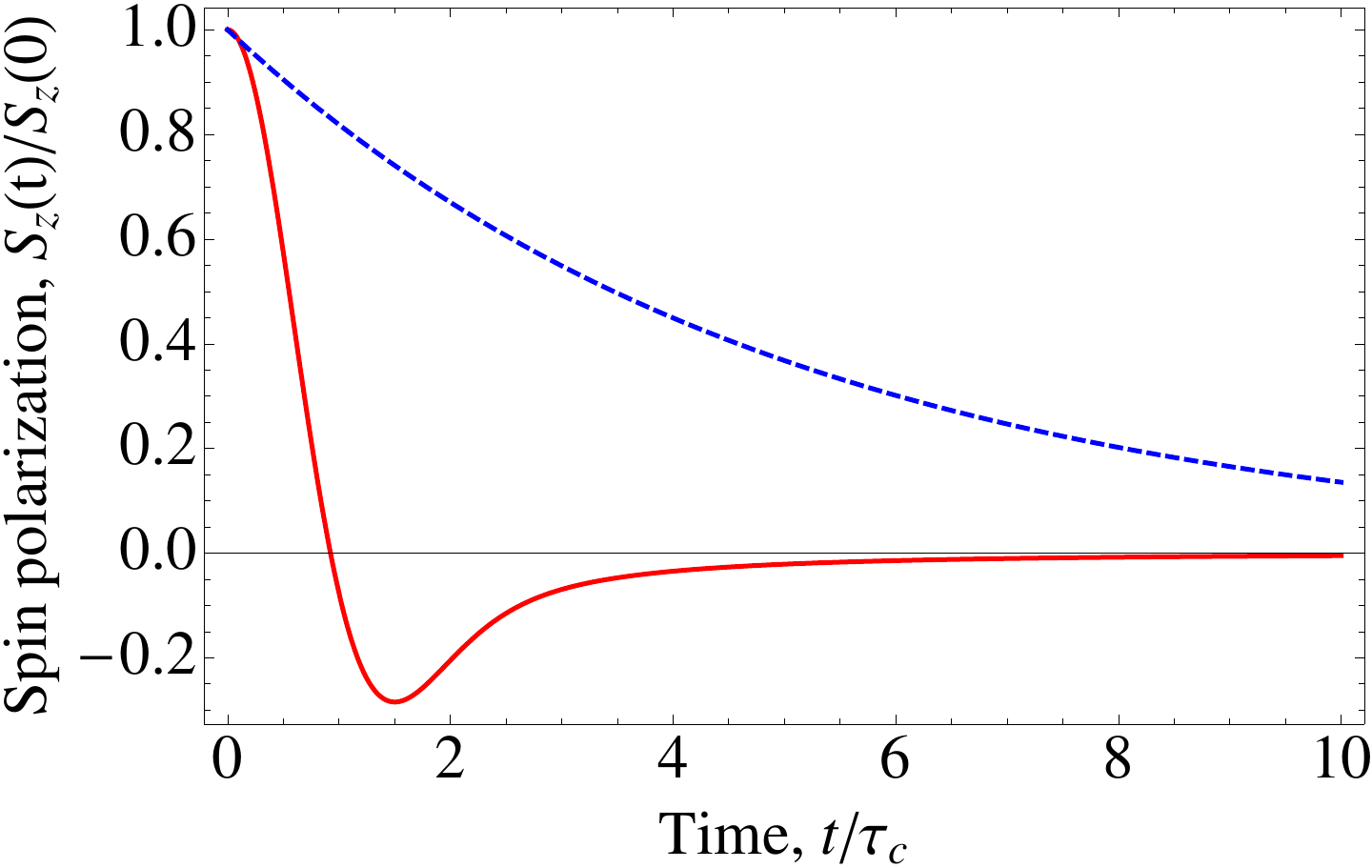}
\caption{Examples of free exciton spin dynamics: Red solid curve shows the spin polarization as a function of time in the absence of scattering {($\tau_2\to \infty$)} calculated after Eq.~\eqref{S:prec} for Boltzmann excitons. Blue dashed curve shows exponential spin relaxation, Eq.~\eqref{S:exp}, for finite scattering time $\tau_{{2}}/\tau_c = 1/5$; $\tau_c$ is related with characteristic spin precession period in the field $\bm \Omega_{\mathbf K}$, see text for details.}\label{fig:Szt}
\end{figure}

In order to illustrate the role of the effective field $\bm \Omega_{\mathbf K}$ in exciton spin/valley depolarization, we consider the hypothetic case where the scattering processes are completely neglected, $\bm Q\{\bm S_{\mathbf K}\}\equiv 0$, but exciton distribution is Boltzmann. This can be achieved  if pump-induced initial distribution is Boltzmann-like or if rapid exciton-exciton collisions quickly brought excitonic ensemble to some internal thermal equilibrium and then become negligible due to the fast recombination process.  Taking into account that the effective field is isotropic in the structure plane we obtain (neglecting the contribution of the small part of the wavevector space within the light cone) for thermalized excitons
\begin{multline}
\label{S:prec}
S_z(t) = \sum_{\mathbf K} S_{z,\mathbf K}(t) = S_z(0) \frac{ \sum_{\mathbf K} \cos{(\Omega_{\mathbf K}t)} n(\varepsilon_{\mathbf K})}{\sum_{\mathbf K} n(\varepsilon_{\mathbf K}) } = \\
S_z(0) \left[1+  \frac{\mathrm i t}{\tau_c} \sqrt{\pi}\erf\left( \frac{\mathrm i t}{\tau_c}\right) \exp{\left(-\frac{t^2}{\tau_c^2}\right)} \right],
\end{multline}
where $\erf(z) = \sqrt{2/\pi} \int_0^z \exp{(-x^2)}dx$ is the error function, and $\tau_c^{-1} = \Omega_{k_T}= \alpha(k_T) k_T^2 /\hbar$, with $k_T = \sqrt{2M k_B T/\hbar^2}$ being thermal wavevector. According to Eq.~\eqref{S:prec} the exciton spin decay is accompanied with one half-period of oscillation, see red solid curve in Fig.~\ref{fig:Szt}. In this case the spin dephasing takes place only due to the spin precession: Excitons with different wave vectors have different precession frequencies resulting in the spin decoherence. The decoherence rate $1/\tau_c$ can be estimated from the LT-splitting at $K \sim k_T$ and scales as $\sqrt{k_B T}$ reflecting the spanning of the LT splitting in the exciton ensemble.  At long times, $t\gg \tau_c$, one has strongly non-exponential decay, $S_z(t)/S_z(0) \approx -\tau_c^2/(2t^2)$. The typical decay time of the in-plane pseudospin components $S_x$, $S_y$ has the same order of magnitude as $\tau_c$. Equation~\eqref{S:prec} shows that even in the absence of scattering the long-range exchange interaction leads to the exciton spin/valley decoherence.

Inclusion of scattering processes makes free exciton spin dynamics qualitatively different. In particular, in the strong scattering regime, where $\Omega_{k_T}\tau_{sc} \ll 1$, where $\tau_{sc}$ is the characteristic scattering time, the spin is lost by Dyakonov-Perel' mechanism~\cite{dyakonov72}. Hence, the spin decay law is exponential
\begin{equation}
\label{S:exp}
S_i(t) = S_i(0) \exp{(-t/\tau_{ii})}, \quad i=x,y,z,
\end{equation}
see blue dashed curve in Fig.~\ref{fig:Szt} and the components of spin relaxation rates tensor are given by~\cite{maialle93,ivchenko05a,PhysRevB.77.165341,glazov2014exciton}
\begin{equation}
\label{tau:s:scatt}
\frac{1}{\tau_{zz}} = \frac{2}{\tau_{xx}} = \frac{2}{\tau_{yy}} = \langle \Omega_{\mathbf K}^2 \tau_2\rangle,
\end{equation}
where the angular brackets denote averaging over the energy distribution and $\tau_2 = \tau_2(\varepsilon_{\mathbf K})$ is the relaxation time of second angular harmonics  of the distribution function. This time describes also the decay of momentum alignment. In derivation of Eq.~\eqref{tau:s:scatt} elastic (or quasi-elastic) processes of exciton-defect and exciton-phonon scattering were taken into account in the collision integral, Eq.~\eqref{Q}, in which case $W_{\mathbf K\mathbf K'} = W_{\mathbf K'\mathbf K} \propto \delta(\varepsilon_{\mathbf K} - \varepsilon_{\mathbf K'})$ and
\begin{equation}
\frac{1}{\tau_2(\varepsilon_{\mathbf K})} = \sum_{\mathbf K'} W_{\mathbf K'\mathbf K}(1-\cos{2\vartheta'}),
\end{equation}
with $\vartheta'$ being the angle between $\mathbf K'$ and $x$-axis. In the strong scattering regime,  $\Omega_{k_T}\tau_2 \ll 1$, the pseudospin rotation angles between the collisions are $\sim \Omega_{k_T} \tau_2 \ll 1$ and spin decoherence results from the random diffusion of the pseudospin over the  sphere. It follows from Eq.~\eqref{tau:s:scatt} that the temperature dependence of the relaxation rates is determined by an interplay of the increase of $\Omega_{\mathbf K}^2\propto k_BT$ with an increase in the temperature and possible energy, and thus temperature dependence of the scattering time $\tau_2$.

Equation~\eqref{tau:s:scatt} shows the anisotropy of exciton spin and valley relaxation. Particularly, circular polarization degree/valley polarization whose relaxation is governed by the rate $1/\tau_{zz}$ decays twice faster compared with the linear polarization degree/inter-valley coherence of excitons whose decay is controlled by smaller rate $1/\tau_{xx} = 1/\tau_{yy}$. This is because the effective field due to the exchange interaction $\bm \Omega_{\mathbf K}$ lies in the structure plane, hence, both $\Omega_x$ and $\Omega_y$ affect circular polarization degree, while linear polarization degree is affected by one component of $\bm \Omega_{\mathbf K}$ only.

Equation~\eqref{kin} is valid for free excitons, while experiments demonstrate that the localized exciton states play an important role in transition metal dichalcogenide monolayers~\cite{Mak:2012qf,Sallen:2012qf,PhysRevB.90.075413,PhysRevB.90.161302}. For completeness, let us discuss the spin decoherence of localized excitons, which can be caused by the fine structure splitting of localized electron-hole pair states provided that the localization potential has an in-plane anisotropy. The latter can be caused by a symmetry reduction induced by the coupling with the substrate, random deformations, anisotropic potential fluctuations, etc. We assume that the exciton is localized as a whole, i.e. its localization lengths $a_{x'}$ and $a_{y'}$ along the main axes of $x'$ and $y'$ of the imperfection exceed by far excitonic effective radius. Hence, in order to obtain an effective Hamiltonian for a localized electron-hole pair one can average $\mathcal H_X$ in Eq.~\eqref{H:bright} over the exciton center of mass wavefunction. As a result, we arrive at two eigenstates linearly polarized along the main axes $x'$ and $y'$ with the anisotropic splitting, cf.~\cite{goupalov98,maialle00}
\begin{multline}
\label{anisotropic}
\Delta E_a = 2 \left(\left\langle \alpha(K){K_{x'}^2}\right \rangle - \left\langle \alpha(K){K_{y'}^2}\right \rangle \right) \\
\sim \frac{\hbar  \Gamma_0}{q} \frac{a_{y'} - a_{x'}}{a_{x'}a_{y'}}.
\end{multline}
Here angular brackets denote quantum-mechanical averaging. The time scale for the spin decoherence of localized excitons is given by $\tau_{c, \rm loc} \sim \hbar /\Delta E_a$. Very recently, the fine structure splittings for localized excitons were observed in 
WSe$_2$ monolayer flakes~\cite{2014arXiv1411.0025S,2014arXiv1411.2449H}. The splittings are about $0.7$~meV and are significantly larger than those in standard semiconductor quantum dots. This enhancement of the $\Delta E_a$ is consistent with stronger excitonic effects, particularly, larger values of the Coulomb energy and of $\Gamma_0$.

\section{Bright exciton spin dynamics. Experiment}\label{sec:exper}

The theory of bright exciton spin dynamics governed by the long-range exchange interaction outlined above has been developed for monolayers of transition metal dichalcogenides in Ref.~\cite{glazov2014exciton}. It has been successfully applied to describe experimental data on MX$_2$ monolayers obtained by time-resolved photoluminescence on MoS$_2$ monolayers~\cite{glazov2014exciton} and time-resolved Kerr rotation on WSe$_2$ monolayers~\cite{PhysRevB.90.161302}. The role of the long-range exchange interaction in the intervalley excitonic transfer has been also confirmed by combination of time-resolved and \emph{cw} polarized photoluminescence spectroscopy of WSe$_2$ monolayers~\cite{2015arXiv150207088Y} and by \emph{cw} photoluminescence spectroscopy on few-layer samples of MoS$_2$~\cite{Kim:2015aa}.

\begin{figure}
\includegraphics[width=\linewidth]{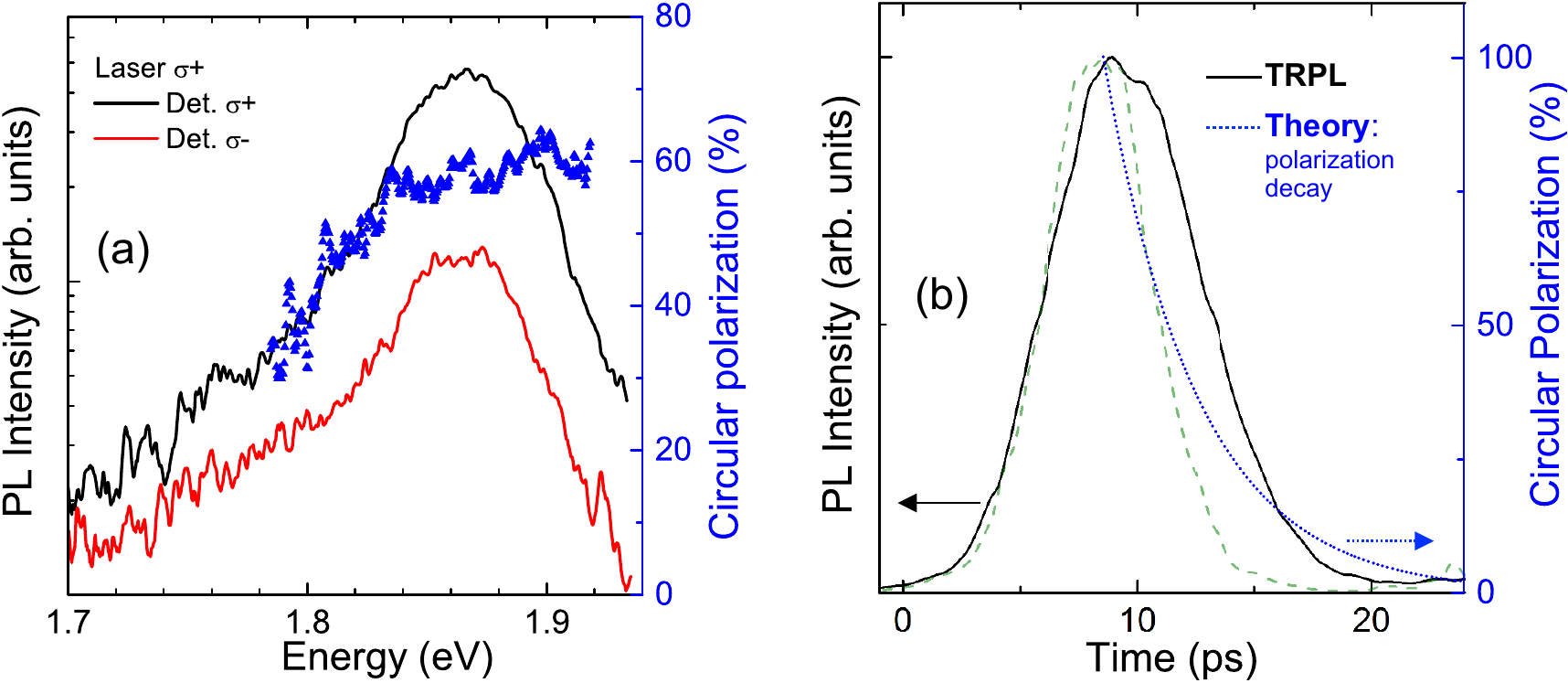}
\caption{ (a) Left axis: Time integrated photoluminescence intensity as a function of emission energy. Right axis: Polarization of photoluminescence emission. (b) Left axis: photoluminescence emission intensity (black line) detected at maximum of A-exciton photoluminescence, $E_{\text{det}}=1.867$~eV as a function of time and laser reference pulse (green dashed line). Right axis: Calculated evolution of photoluminescence polarization after Eq.~\eqref{S:exp} using $\tau_{zz}\approx 4$~ps (blue dotted line).  $T=4$~K, $E_{\text{exc}}$=1.965~eV, $P_{\text{exc}} =5~\mu$W$/\mu$m$^2$. From Ref.~\cite{glazov2014exciton}}\label{fig:TRPL}
\end{figure} 

Figure~\ref{fig:TRPL} presents the results of photoluminescence experiments carried out in Ref.~\cite{glazov2014exciton} on MoS$_2$ monolayer deposited on the SiO$_2$/Si substrate. These experiments as well as the experiments in Ref.~\cite{PhysRevB.90.161302} and described below have been performed with the samples in vacuum. Both photoluminescence intensity and its circular polarization degree (under circularly polarized excitation) were recorded as a function of emission energy. As a simplest possible model, we assume that the stationary, i.e. time integrated, polarization is determined by the initially created polarization $P_0$, the lifetime of the electron-hole pair $\tau$ and the polarization decay time $\tau_{s}$ as~\cite{opt_or_book}:
\begin{equation}
\label{Pc}
P_c=\frac{P_0}{1+\tau/\tau_{s}}.
\end{equation}
In Fig.~\ref{fig:TRPL}(a) an average, time-integrated photoluminescence polarization of $P_c\approx 60\%$ in the emission energy range $1.82\ldots 1.91$~eV is observed. The emission time measured in time-resolved photoluminescence experiments is $\tau \simeq 4.5$~ps, extracted from Fig.~\ref{fig:TRPL}(b) in agreement with earlier measurements~\cite{PhysRevLett.112.047401}. For the studied MoS$_2$ it was not possible to conclude if the measured emission time is a radiative lifetime or it is limited by non-radiative processes. For $P_0=100\%$ we find an  estimate of $\tau_{s}\simeq 7$~ps. This value is in reasonable agreement with crude theoretical estimates of $\tau_{zz} {\sim} 4$~ps made for this sample under the assumption that the spread of excitons in the energy space is limited by the collisional broadening, $\sim \hbar/\tau_2$, rather than by the kinetic energy distribution~\cite{maialle93,glazov2014exciton}. Note, that in this case the criteria of applicability of kinetic Eq.~\eqref{kin} are not fulfilled, which results in inaccuracy of $\tau_{zz}$ estimation. It is worth stressing, that even with a value of $\tau_{zz}$ in the picosecond range a high polarization can be obtained in time integrated measurements. This is because the photoluminescence decay time $\tau$ is also ultrashort. It is demonstrated in Fig.~\ref{fig:TRPL}(b), which allows to compare the theoretically predicted polarization decay and the measured photoluminescence intensity decay.

\begin{figure}[t]
\includegraphics[width=\linewidth]{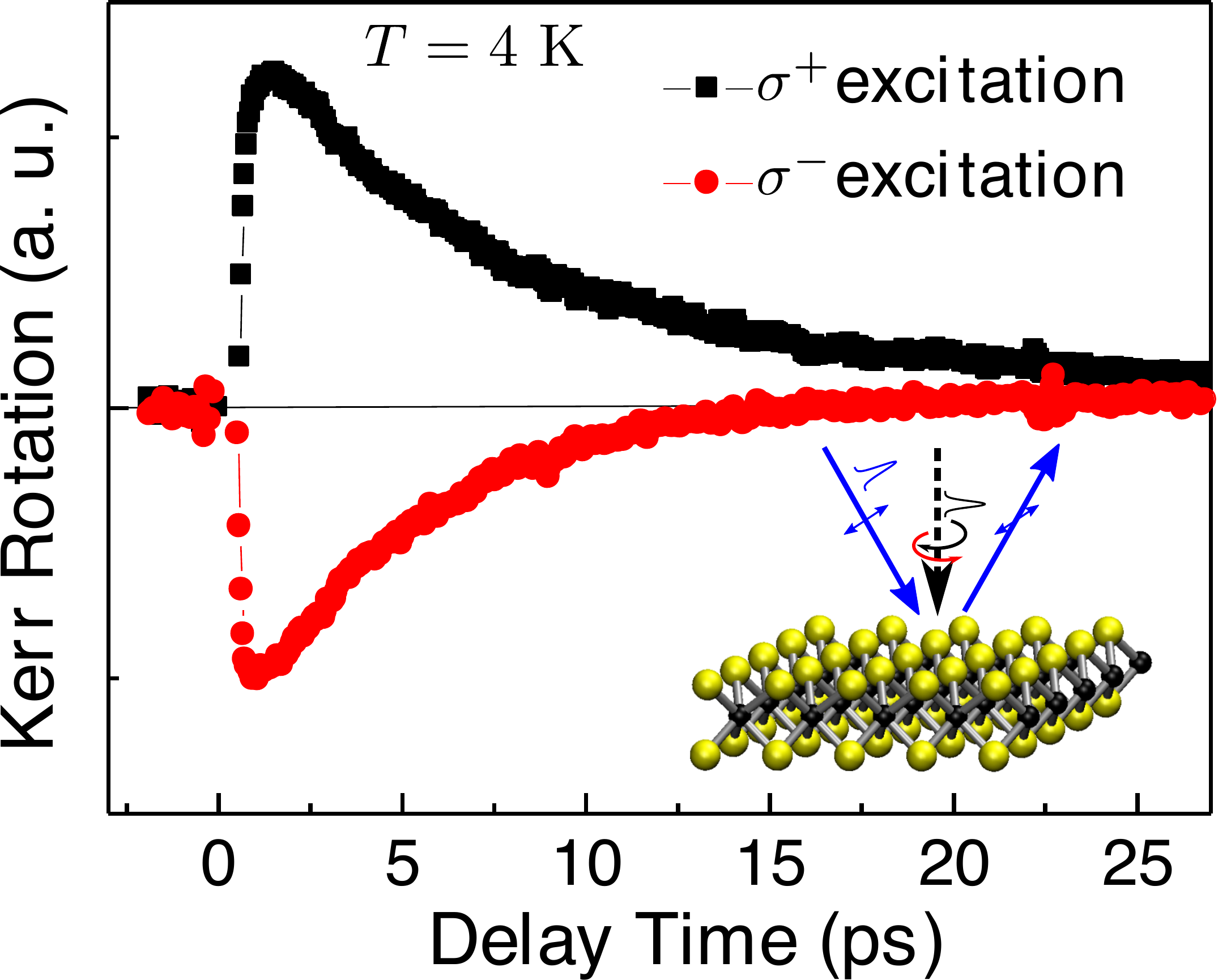}
\caption{ Kerr rotation dynamics at $T=4$~K for $\sigma^+$ and $\sigma^-$ pump beam. The laser excitation energy is $E_l=1.735$~eV. Inset shows sketch of time resolved Kerr rotation effect. Data from Ref.~\cite{PhysRevB.90.161302}.}\label{fig:fig2}
\end{figure} 

Time-resolved Kerr rotation technique serves as an alternative and highly sensitive tool to study spin dynamics, see Refs.~\cite{dyakonov_book,glazov:review} for review. In this method the spin-Kerr effect, i.e. polarization plane rotation angle of the linearly polarized probe pulse, $\theta(\Delta t)$, is measured as a function of the time delay, $\Delta t$, between the probe and preceding circularly polarized pump pulse, see inset in Fig.~\ref{fig:fig2}. The pump pulse creates spin/valley polarization of excitons, which results in polarization-dependent modification of the parameters $\omega_0$, $\Gamma_0$ and $\Gamma$ in the MX$_2$ reflection coefficient, $r(\omega)$, Eq.~\eqref{normal} due to nonlinearities in the system~\cite{glazov:review}. Thus, the probe beam monitors imbalance of $\sigma^+$ and $\sigma^-$ excitons in the system. 

Such measurements were carried out in Ref.~\cite{PhysRevB.90.161302} on high quality WSe$_2$ monolayers where the trion and exciton lines in photoluminescence are separated by $\sim 30$~meV and the exciton line is stable up to the room temperature. The latter makes it possible to experimentally address temperature dependence of exciton spin dynamics, unaccessible in conventional semiconductors. Additional advantage of the Kerr rotation technique comes from the fact that the limited time resolution of photoluminescence spectroscopy does not allow one to address spin dynamics at elevated temperatures where spin relaxation time decreases~\cite{PhysRevLett.112.047401,PhysRevB.90.075413}. Kerr rotation dynamics has higher time resolution ($\sim$100 fs) and makes it possible to apply a strictly resonant excitation of the exciton.

\begin{figure}[t]
\includegraphics[width=0.5\textwidth]{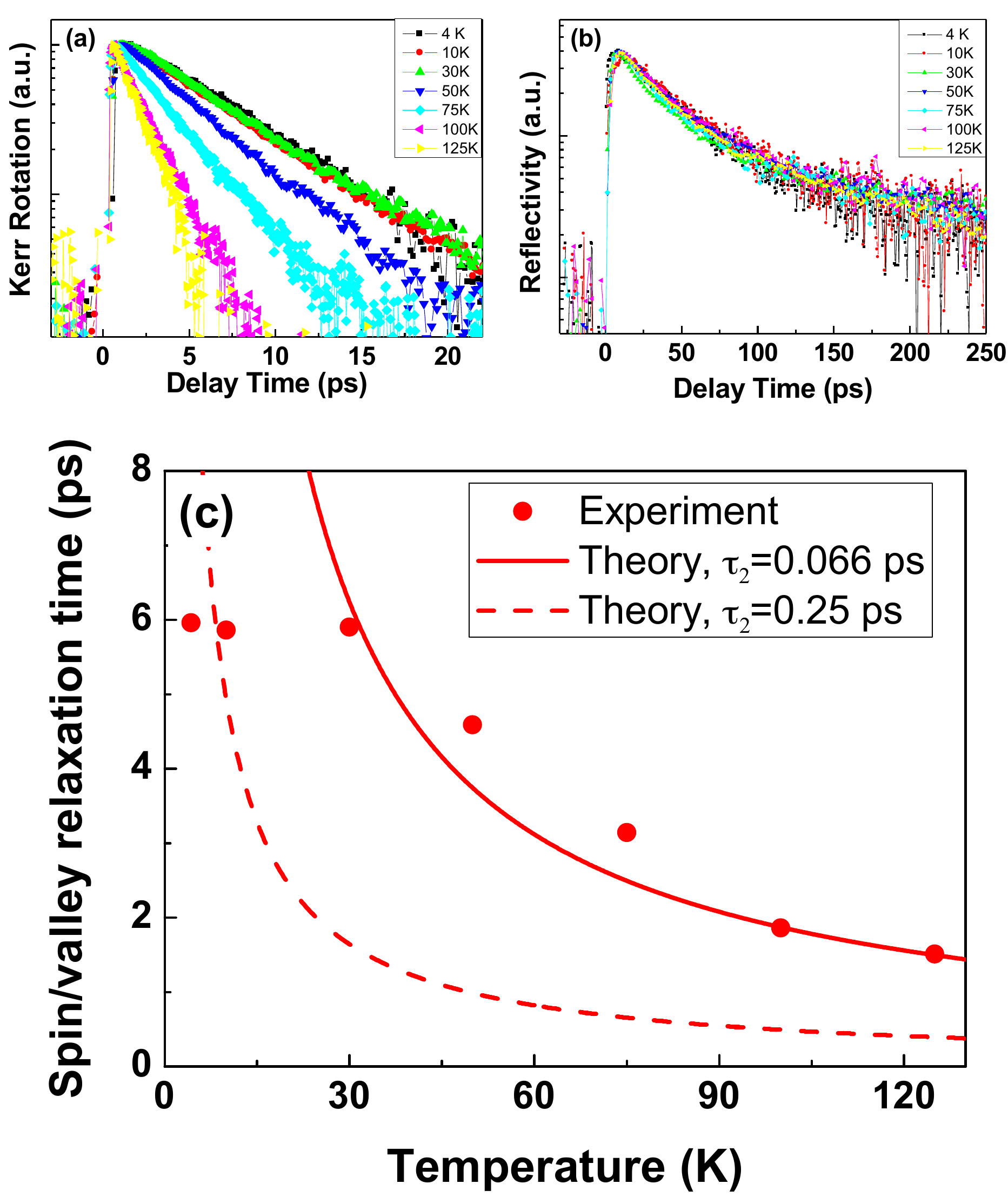}
\caption{ (a) Kerr rotation dynamics after a $\sigma^+$ polarized pump 
pulse for different lattice temperatures;
(b)	Transient reflectivity dynamics for different temperatures. The laser excitation energies are identical to the ones used in (a), see text;
(c)	Temperature dependence of the measured (symbol) and calculated after Eq.~\eqref{tau:s:scatt} (solid and dashed lines) exciton valley polarization relaxation time, see text for details.
 From Ref.~\cite{PhysRevB.90.161302}.}\label{fig:fig3}
\end{figure} 
 
Figure~\ref{fig:fig2}  shows the Kerr rotation dynamics, $\theta(\Delta t)$, measured at 4 K for both $\sigma^+$ and $\sigma^-$ polarized pump pulses. The pump energy $E_{l}=  1.735$~eV is set to the maximum of the Kerr signal, which is very close to the neutral exciton transition identified in the photoluminescence spectra. The observed sign reversal of the Kerr signal in Fig.~\ref{fig:fig2} at the reversal of pump helicity is a consequence of the selective optical initialization of the $\bm K_+$ and $\bm K_-$ valley, respectively. The transient reflectivity measured at the same conditions using linearly cross-polarized pump and probe pulses is shown in Fig.~\ref{fig:fig3}(b). It is seen that the reflectivity decay time is about ten times longer than the one observed in time-resolved Kerr rotation. Therefore the mono-exponential decay time $\tau_s=(6 \pm 0.1)$~ps of the Kerr rotation dynamics at $T=4$~K in Fig.~\ref{fig:fig2}, probes directly the fast exciton valley depolarization. The value of $\tau_s$ is close to that measured by time-resolved photoluminescence on MoS$_2$ samples, see above.

Figure~\ref{fig:fig3}(a) displays the variation of the spin-Kerr effect dynamics as a function of the temperature. For $T\gtrsim 30$~K, a clear decrease of the exciton spin/valley polarization decay time $\tau_s$ is observed down to 1.5 ps at $T=125$~K. At the same conditions reflection coefficient decay is much longer, see Fig.~\ref{fig:fig3}(b). The excitation power dependences of the exciton dynamics have been also investigated. In the studied power range corresponding to variation of exciton density from $n_x\sim 1.5 \times 10^{11}$ to $10^{12}$~cm$^{-2}$, both the Kerr rotation and reflectivity dynamics do not depend on the  photo-generated exciton density within the experimental accuracy. This demonstrates that the exciton-exciton interactions play a minor role in exciton valley dynamics presented in Figs.~\ref{fig:fig2} and \ref{fig:fig3} over this density range. Note, that in Ref.~\cite{2015arXiv150207088Y} much higher excitation intensities were applied to WSe$_2$ monolayers resulting in the exciton densities up to $n_x \gtrsim10^{14}$~cm$^{-2}$. In this case the effect of exciton density on circular polarization of emission was observed. Particularly, Ref.~\cite{2015arXiv150207088Y}  reports an increase of the circular polarization degree with an increase of exciton density up to $5\times 10^{13}$~cm$^{-2}$ and a decrease of the polarization degree with further increase in $n_x$.

Figure~\ref{fig:fig3}(c) shows the experimentally measured in Ref.~\cite{PhysRevB.90.161302} exciton polarization decay times and theoretical calculations carried out after Eqs.~\eqref{alpha1} and \eqref{tau:s:scatt} for $n=\sqrt{10}$ and $\Gamma_0 = 0.16$~ps$^{-1}$. Qualitatively, the drop of the exciton spin/valley relaxation time when the temperature increases can be well explained by the increase with the temperature of the precession frequency $\bm \Omega_{\mathbf K}\propto K$, which makes spin precession and decoherence faster, Eq.~\eqref{tau:s:scatt}. The solid and dashed lines in Fig.~\ref{fig:fig3}(c) correspond to the calculated exciton spin/valley relaxation time for  two different scattering times $\tau_2=0.066$~ps and $\tau_2=0.25$~ps, respectively. 

The choice of the scattering time  values in the calculations of Ref.~\cite{PhysRevB.90.161302} was motivated by the following reasons. The latter (longer) value of $\tau_2$ corresponds to exciton energy uncertainty equivalent to $30$~K corresponding to Ioffe-Regel criterion of delocalization where the product of the thermal wavevector $k_T$ and the mean free path $l$ is on the order of $1$. However, a much better agreement between the calculated and measured exciton relaxation times in Fig.~\ref{fig:fig3}(c) for $T> 30$~K  is observed for the smaller value of the  scattering time $\tau_2=0.066$~ps, but formal criterion of kinetic equation is fulfilled in this case only for high temperatures, $T\gtrsim 100$~K. The possible origins of the scattering are: (i) short-range defects, (ii) Coulomb scattering with resident electrons. In the latter case for resident electron density $n_e \sim 10^{12}$~cm$^{-2}$ the exchange exciton-electron~\cite{tarasenko98} scattering yields $\tau_{{2}} \sim 10^{-13}$~s. 

At low temperatures, $4<T<30$~K, the measured exciton spin relaxation time is almost temperature independent. This behavior could be either caused by (i) the very strong scattering in a regime where $k_BT$ is smaller than the collision broadening leading to a temperature independent spin relaxation time, see Refs.~\cite{maialle93,glazov2014exciton} and discussion of experiment on MoS$_2$ above, or (ii) by a localized character of the exciton below 30~K. Valley dynamics of excitons in WSe$_2$ has been also addressed by photoluminescence spectroscopy~\cite{2015arXiv150207088Y} and by Kerr rotation spectroscopy~\cite{china2}. The origin of discrepancy of the results of Refs.~\cite{2015arXiv150207088Y} and \cite{china2} as well as of Ref.~\cite{PhysRevB.90.161302} is not clear and is probably related to different experimental methods.

\begin{figure}[t]
\includegraphics[width=\linewidth]{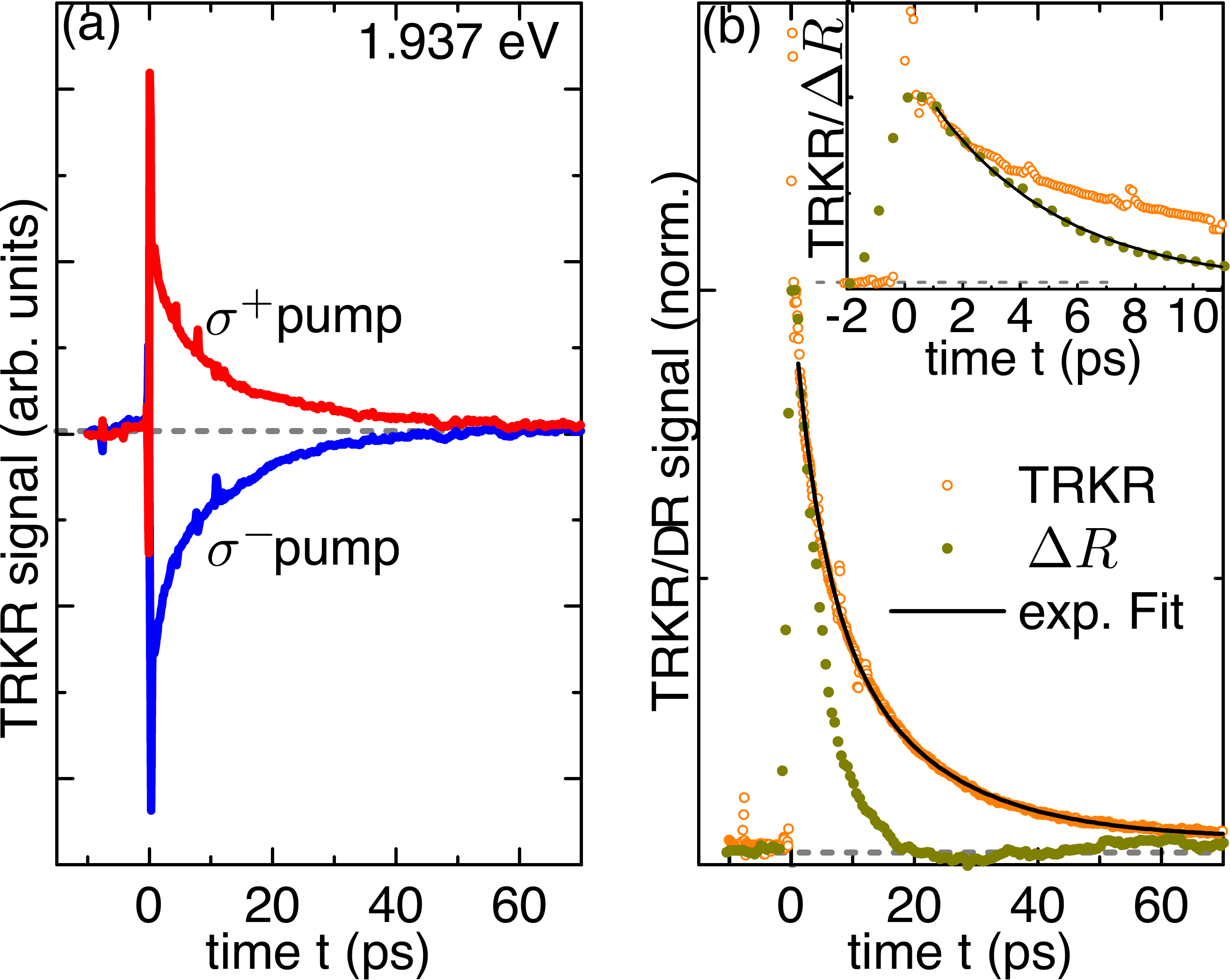}
\caption{(a) TRKR traces measured on the MoS$_2$ flake for different helicities of the pump beam, with a laser excitation energy of 1.937 eV. (b) Normalized time resolved Kerr rotation (TRKR) and transient reflectivity traces measured on the same flake with a laser excitation energy of 1.937 eV. The solid line indicates a biexponential fit to the time resolved Kerr rotation trace. The inset shows a high-resolution plot of the data, The solid line indicates a monoexponential fit to the transient reflectivity trace. From Ref.~\cite{2014arXiv1404.7674P}.}\label{fig:Korn}
\end{figure} 

In experiments performed in Ref.~\cite{PhysRevB.90.161302} no long-living spin/valley polarization was detected: The Kerr rotation signal is absent for the pump-probe delays $\Delta t\gg 25$~ps, where photoinduced reflectivity vanishes. It indicates absence of polarization transfer from excitons to resident carriers in WSe$_2$, unlike similar time-resolved Kerr experiments performed in III-V or II-VI semiconductor nanosystems, see Refs.~\cite{glazov:review,dyakonov_book} for review. By contrast, experiments performed on MoS$_2$ monolayer flakes in Ref.~\cite{2014arXiv1404.7674P} indicate valley polarization of resident carriers. Figure~\ref{fig:Korn}(a) demonstrates Kerr rotation signal as a function of the pump-probe delay for $\sigma^+$ and $\sigma^-$ helicities of the pump pulse. Like the data shown in Fig.~\ref{fig:fig2} the signal flips its sign with reversal of radiation helicity evidencing selective population of $\bm K_\pm$ valleys. However, unlike data shown in Fig.~\ref{fig:fig3}, the transient reflectivity measured in Ref.~\cite{2014arXiv1404.7674P} on MoS$_2$ flake decays on the timescale of $\sim 4.5$~ps, while the Kerr rotation signal demonstrates bi-exponential behavior, see Fig.~\ref{fig:Korn}(b). The analysis carried out in Ref.~\cite{2014arXiv1404.7674P} shows that this decay contains two components, the fast one with the same decay time as that of the reflectivity, and the slow one with the longer decay time of 17~ps at $T=4.5$~K. This long decay time is attributed in Ref.~\cite{2014arXiv1404.7674P}  to the spin/valley polarization of resident electrons in the sample. In this case time resolved Kerr rotation technique provides an access to the resident electron spin dynamics in transition metal dichalcogenide sample.

\section{Outlook}\label{sec:out}

The understanding of spin and valley dynamics of excitons in transition metal dichalcogenides has rapidly evolved over the last three years. Although this area of research started only in 2012, a number of solid experimental facts has been already established and conclusive theoretical models have been developed. Particularly, it has been demonstrated that the spin and valley dynamics of bright exciton doublet in MX$_2$ monolayers is governed by the long-range exchange interaction between an electron and a hole forming an exciton. This interaction acts, in the pseudospin picture, as an effective magnetic field coupling the circularly polarized states of moving exciton. The long-range exchange interaction can also be responsible for the localized excitons fine structure and their spin decoherence.

Still, further progress in needed before a complete understanding of multifaceted excitonic spin dynamics in MX$_2$ monolayers can be reached. Despite strong theoretical and experimental progress the full picture of the spin dynamics of charged excitons is still absent. It is still not clear why no optical orientation of excitons is observed in high-quality MoSe$_2$ monolayers~\cite{:/content/aip/journal/apl/106/11/10.1063/1.4916089}. Moreover, the magnetooptical studies of excitons and trions in transition metal dichalcogenides such as WSe$_2$ and MoSe$_2$ reveal large exciton $g$-factors in magnetic field perpendicular to the sample plane, $|g_x|\sim 4$~\cite{PhysRevLett.114.037401,Li:2014a,Srivastava:2015a,Aivazian:2015a,2015arXiv150304105W}. These observations are very surprising because the two-band $\bm k \cdot \bm p$ model, Eq.~\eqref{Hkp} used here, results in $g_x \equiv 0$~\cite{2015arXiv150304105W}. Therefore further experimental and theoretical studies are needed to elaborate the band structure description of transition metal dichalcogenides monolayers.

%
%
%

\begin{acknowledgement}
We acknowledge partial funding from Programs of Russian Academy of Sciences, RFBR, RF President grants MD-5726.2015.2 and NSh-1085.2014.2 and Dynasty Foundation -- ICFPM, ERC Grant No. 306719, ANR MoS2ValleyControl, Programme Investissements d'Avenir ANR-11-IDEX-0002-02, reference ANR-10-LABX-0037-NEXT.
\end{acknowledgement}

%
\bibliographystyle{pss}

\providecommand{\WileyBibTextsc}{}
\let\textsc\WileyBibTextsc
\providecommand{\othercit}{}
\providecommand{\jr}[1]{#1}
\providecommand{\etal}{~et~al.}

%

\end{document}